\documentclass[12pt,showpacs,english]{revtex4}
\usepackage[T1]{fontenc}
\usepackage[latin1]{inputenc}
\usepackage{babel}
\usepackage{graphicx}
\usepackage{setspace}
\begin{document}
\def\sl#1{\slash{\hspace{-0.2 truecm}#1}}
\def\beqn{\begin{eqnarray}}
\def\eeqn{\end{eqnarray}}
\def\nn{\nonumber}
%
%
\title{Resonance estimates for single spin asymmetries in elastic electron-nucleon scattering}
\author{B. Pasquini}
\affiliation{Dipartimento di Fisica Nucleare e Teorica, Universit\`a degli
Studi di Pavia and INFN, Sezione di Pavia, Pavia, Italy}
\affiliation{ECT$^*$, Villazzano (Trento), Italy}
\author{M. Vanderhaeghen}
\affiliation{Thomas Jefferson National Accelerator Facility,
Newport News, VA 23606, USA}
\affiliation{Department of Physics, College of William and Mary,
Williamsburg, VA 23187, USA}
\date{\today}
\begin{abstract}
We discuss the target and beam normal spin asymmetries in 
elastic electron-nucleon scattering which depend on the imaginary part 
of two-photon exchange processes between electron and nucleon. 
We express this imaginary part as a phase space integral over the doubly 
virtual Compton scattering tensor on the nucleon. 
We use unitarity to model the doubly virtual Compton scattering 
tensor in the resonance region in terms 
of $\gamma^* N \to \pi N$ electroabsorption amplitudes. Taking those 
amplitudes from a phenomenological analysis 
of pion electroproduction observables, we present results for beam and 
target normal single spin asymmetries 
for elastic electron-nucleon scattering 
for beam energies below 1 GeV and in the 1-3 GeV 
region, where several experiments are performed or are in progress. 
\end{abstract}
\maketitle 
\section{Introduction}

Elastic electron-nucleon scattering in the one-photon exchange approximation 
is a time-honoured tool to access information on the structure of hadrons. 
Experiments with increasing precision have become possible in recent years, 
mainly triggered by new techniques to perform polarization experiments 
at the electron scattering facilities. This has allowed to reach a new 
frontier in the measurement of hadron structure quantities, such as 
its electroweak form factors, parity violating effects, 
nucleon polarizabilities, $N \to \Delta$ transition form factors, or 
the measurement of spin dependent structure functions, to name a few. 
For example, experiments using 
polarized electron beams and measuring the ratio of the recoil nucleon 
in-plane polarization components have profoundly extended our understanding 
of the nucleon electromagnetic form factors. 
For the proton, such polarization experiments which access 
the ratio $G_{E p} / G_{M p}$ of the proton's electric ( $G_{E p}$ ) to 
magnetic ( $G_{M p}$ ) form factors have been performed 
out to a momentum transfer $Q^2$ of 5.6 GeV$^2$~\cite{Jones00,Gayou02} . 
It came as a surprise that these experiments 
extracted a ratio of $G_{E p} / G_{M p}$ which is clearly at variance with 
unpolarized measurements \cite{Slac94,Chr04,Arr03} 
using the Rosenbluth separation technique. 
\newline
\indent
The understanding of this puzzle has generated a lot of activity recently, 
and is a prerequisite to use electron scattering as a precision tool. 
It has been suggested on general grounds 
in Ref.~\cite{GV03} that this puzzle may be 
explained by a two-photon exchange amplitude  
of the level of a few percent. 
The resulting failure of the one-photon exchange approximation to 
unpolarized elastic electron-nucleon scattering can be understood 
from the observation that $G_{M p}$ and $G_{E p}$ 
enter quadratically in the unpolarized cross section. 
It turns out that $G_{E p}^2$ may become a small quantity 
compared with $G_{M p}^2$, 
and is further suppressed by a kinematical factor $\sim 1/Q^2$. 
Therefore, it becomes increasingly difficult 
to extract this term at larger momentum transfer. 
Already at moderate momentum transfers, the weight of the term 
proportional to $G_E^2$ drops at the 1 \% level and one may expect that  
correction terms due to two-photon exchange become competitive and 
eventually dominate over the $G_E^2$ term.    
The polarization transfer method on the other hand is much less affected  
because it directly measures the ratio of $G_{E p} / G_{M p}$ , 
i.e. depends linearly on $G_E$. 
\newline
\indent
Recently, several model calculations of the $2 \gamma$ exchange 
amplitude have been performed. In Ref.~\cite{BMT03}, 
a calculation of the $2 \gamma$ exchange when the 
hadronic intermediate state is a nucleon was performed. 
It found that the $2 \gamma$ exchange correction with 
intermediate nucleon can partially resolve the discrepancy between the two 
experimental techniques. 
Recently, the $2 \gamma$ exchange contribution to elastic 
electron-nucleon scattering has been estimated 
at large momentum transfer~\cite{YCC04}, 
through the scattering off a parton in a proton by relating 
the process on the nucleon to the generalized parton distributions. 
This calculation found that the $2 \gamma$ exchange contribution is 
indeed able to quantitatively resolve the existing discrepancy between 
Rosenbluth and polarization transfer experiments. 
\newline
\indent 
To push the precision frontier further in electron scattering, one needs 
a good control of $2 \gamma$ exchange mechanisms 
and needs to understand how they may or may not affect 
different observables. This justifies a systematic study of such 
$2 \gamma$ exchange effects, both theoretically and experimentally.  
The real (dispersive) part of the $2 \gamma$ exchange amplitude 
can be accessed through the difference between 
elastic electron and positron scattering off a nucleon.  
The imaginary (absorptive) part of the $2 \gamma$ exchange amplitude 
on the other hand can be accessed through a single spin asymmetry (SSA) in 
elastic electron-nucleon scattering, when either the target or beam spin 
are polarized normal to the scattering plane, as has been discussed some time 
ago in Ref.~\cite{RKR71}. As time reversal invariance forces 
this SSA to vanish for one-photon exchange, it 
is of order $\alpha = e^2 / (4 \pi) \simeq 1/ 137$. Furthermore, to polarize   
an ultra-relativistic particle in the direction 
normal to its momentum involves 
a suppression factor $m / E$ (with $m$ the mass and $E$ the 
energy of the particle), 
which typically is of order $10^{-4} - 10^{-3}$ when the electron 
beam energy is in the 1 GeV range. Therefore, the 
resulting target normal SSA can be expected to be of order $10^{-2}$, 
whereas the beam normal SSA is of order $10^{-6} - 10^{-5}$.
A measurement of such small asymmetries is quite demanding experimentally. 
However, in the case of a polarized lepton beam, asymmetries of the order ppm 
are currently accessible in parity violation (PV) elastic electron-nucleon 
scattering experiments. 
The parity violating asymmetry involves a beam spin polarized 
along its momentum. However the SSA for an electron 
beam spin normal to the scattering plane can also be measured using the 
same experimental set-ups. 
First measurements of this beam normal SSA at beam energies below 1 GeV 
have yielded values around 10 ppm~\cite{Wells01,Maas03}. 
At higher beam energies, the beam normal SSA can also be measured in upcoming  
PV elastic electron-nucleon scattering experiments~\cite{Happex,G0,E158}. 
\newline
\indent
First estimates of the target normal SSA in elastic electron-nucleon 
scattering have been performed in
Refs.~\cite{RKR71,RKR73}. In those works, the $2 \gamma$ exchange 
with nucleon intermediate state (so-called elastic or nucleon 
pole contribution) has been calculated, and the inelastic contribution 
has been estimated in a very forward angle approximation. 
Estimates within this approximation have also been reported for the 
beam normal SSA in Ref.~\cite{AAM02}. 
Recently, the general formalism for elastic electron-nucleon scattering with 
lepton helicity flip, which is needed to describe the beam normal SSA, 
has been developed in Ref.~\cite{GGV04}. 
Furthermore, the beam normal SSA has also been estimated at large 
momentum transfers $Q^2$ in Ref.~\cite{GGV04} using a parton model, 
which was found crucial \cite{YCC04} to interpret the 
results from unpolarized electron-nucleon elastic scattering, 
as discussed before. 
In the handbag model of Refs.~\cite{YCC04,GGV04}, 
the corresponding $2 \gamma$ exchange amplitude 
has been expressed in terms of generalized parton distributions, and 
the real and imaginary part of the 
$2 \gamma$ exchange amplitude are related through a dispersion relation. 
Hence in the partonic regime, 
a direct comparison of the imaginary part with experiment 
can provide a very valuable cross-check 
on the calculated result for the real part.  
\newline
\indent
To use the elastic electron-nucleon scattering at low momentum 
transfer as a high precision tool, such as 
in present day PV experiments, one may also want to quantify the 
$2 \gamma$ exchange amplitude. To this aim, one may 
envisage a dispersion formalism for the elastic electron-nucleon scattering 
amplitudes, as has been discussed some time ago 
in the literature~\cite{BPB87}. 
To develop this formalism, the necessary first step 
is a precise knowledge of the imaginary part of the two-photon exchange 
amplitude, which enters in both the beam and target normal SSA. 
The study of this imaginary part of the $2 \gamma$ exchange is 
the subject of this paper. 
Using unitarity, one can relate the imaginary part of the $2 \gamma$  
amplitude to the electroabsorption amplitudes on a nucleon. 
When measuring the imaginary part of the elastic electron-nucleon 
amplitude through a normal SSA at sufficiently low energies,  
below or around two-pion production threshold, 
one is in a regime where these electroproduction amplitudes are relatively 
well known using pion electroproduction experiments as input. 
One strategy is therefore to investigate this new tool 
of beam and target normal SSA first in the 
region where one has a good first knowledge of the imaginary part 
of the $2 \gamma$ exchange.  
As both photons in the $2 \gamma$ exchange process are 
virtual and integrated over, an observable such as the beam or target normal 
SSA is sensitive to the electroproduction amplitudes 
on the nucleon for a range of photon virtualities. 
This may provide information on resonance transition 
form factors complementary to the information obtained from current 
pion electroproduction experiments.  
\newline
\indent
Finally, by understanding the $2 \gamma$ exchange contributions 
for the case of electromagnetic 
electron-nucleon scattering, one may extend this calculation  
to electroweak processes, where the $\gamma Z$ and $W^+ W^-$ box diagrams 
are in several cases the leading unknown contributions 
entering in electroweak precision experiments.
\newline
\indent
We start by briefly reviewing 
the elastic electron-nucleon scattering formalism 
beyond the one-photon exchange approximation in Section~\ref{sec:en}, 
and discuss the target and beam normal spin asymmetries in 
Section~\ref{sec:ssa}. Subsequently, we study the imaginary part 
of the two-photon exchange amplitudes in Section~\ref{sec:impart}. 
We express this imaginary part as a phase space integral over the doubly 
virtual Compton scattering tensor on the nucleon. 
In Section~\ref{sec:models}, we use unitarity to model the doubly 
virtual Compton scattering tensor in the resonance region in terms 
of $\gamma^* N \to \pi N$ electroabsorption amplitudes. We take those 
amplitudes from a state-of-the-art phenomenological analysis 
(MAID \cite{Dre99}) of pion electroproduction observables. 
In Section~\ref{sec:results}, we show our results for beam and 
target normal SSA for beam energies below 1 GeV and in the 1-3 GeV 
region, where several experiments at MIT-Bates, MAMI and 
Jefferson Lab (JLab) are performed or in progress. 
Our conclusions and an outlook are given in Section~\ref{sec:concl}.

\section{Elastic electron-nucleon scattering beyond the one-photon
exchange approximation}
\label{sec:en}

In this section, we briefly review the elastic electron-nucleon scattering 
formalism beyond the one-photon exchange approximation, as has been developed 
recently in Refs.~\cite{GV03,GGV04}.
For the kinematics of elastic electron-nucleon
scattering~: 
\begin{equation}
\label{Eq:intro.2}
e^-(k)+N(p)\rightarrow e^-(k')+N(p'),
\end{equation}
we adopt the usual definitions~:
\begin{equation}
\label{Eq:intro.3}
P=\frac{p+p'}{2},\, K=\frac{k+k'}{2},\, q=k-k'=p'-p,
\end{equation}
and choose 
\begin{equation}
\label{eq:qsqrnu}
Q^{2}=-q^{2},\, \nu =K.P ,
\end{equation}
as the independent invariants of the scattering. 
The invariant $\nu$ is related to 
the polarization parameter $\varepsilon$ of the virtual photon, 
which can be expressed as (neglecting the electron mass)~:
\begin{equation}
\label{eq:eps}
\varepsilon =\frac{\nu ^{2}-M^{4}\tau (1+\tau )}{\nu ^{2}+M^{4}\tau (1+\tau )},
\end{equation}
where $M$ is the nucleon mass.
\newline
\indent
For a theory which respects Lorentz, parity and charge
conjugation invariance, the general amplitude for elastic scattering
of two spin 1/2 particles can be expressed by 
6 independent helicity amplitudes or equivalently by six invariant amplitudes. 
The total amplitude can be decomposed in general 
in terms of a lepton spin non-flip and spin flip part~:
\begin{eqnarray}
T \;=\; T_{non-flip} \;+\; T_{flip} .
\label{eq:tampl}
\end{eqnarray}
The non-flip amplitude which conserves the helicity of the electron 
(in the limit $m_e = 0$) depends upon 3 invariant amplitudes, and has  
been parametrized in Ref.~\cite{GV03} as~:
\begin{eqnarray}
\label{eq:non-flip}
T_{non-flip} \,&=&\, 
\frac{e^{2}}{Q^{2}} \, \bar{u}(k')\gamma _{\mu }u(k)\, \cdot \,
\bar{u}(p')\left( \tilde{G}_{M}\, \gamma ^{\mu }
-\tilde{F}_{2}\frac{P^{\mu }}{M}
+\tilde{F}_{3}\frac{\gamma .KP^{\mu }}{M^{2}}\right) u(p).
\end{eqnarray}
The amplitude which flips the electron helicity (i.e. is of the
order of the mass of the electron, $m_e$), depends on 3 additional invariants 
which have been introduced in Ref.~\cite{GGV04} as~:
\begin{eqnarray}
\label{eq:flip}
\hspace{-0.4cm}
T_{flip} \,&=&\, \frac{m_e}{M} \,
\frac{e^{2}}{Q^{2}} \left[ \, \bar{u}(k') u(k)\, \cdot \,
\bar{u}(p')\left( \tilde{F}_{4}\, +\tilde{F}_{5}\frac{\gamma . K}{M} 
\right) u(p) \,+\, 
\tilde{F}_{6}\,\bar{u}(k') \gamma_5 u(k)\, \cdot \,\bar{u}(p') \gamma_5 u(p)
\right] .
\end{eqnarray}
In Eqs.~(\ref{eq:non-flip},\ref{eq:flip}), 
$\tilde{G}_{M}$, $\tilde{F}_{2}$, $\tilde{F}_{3}$,
$\tilde{F}_{4}$, $\tilde{F}_{5}$, $\tilde{F}_{6}$ are 
complex functions of $ \nu $ and $ Q^{2} $, and 
the factor $ e^{2}/Q^{2} $ has been introduced for convenience. 
Furthermore in Eq.~(\ref{eq:flip}), we extracted an explicit factor 
$m_e / M$ out of the amplitudes, which reflects the fact that 
for a vector interaction (such as in QED), 
the electron helicity flip amplitude 
vanishes when $m_e \to 0$. 
In the Born approximation, one obtains~:
\begin{eqnarray}
\label{eq:born}
\tilde{G}_{M}^{Born}(\nu ,Q^{2}) \,&=&\, G_{M}(Q^{2}),   \nonumber\\
\tilde{F}_{2}^{Born}(\nu ,Q^{2}) \,&=&\, F_{2}(Q^{2}),   \nonumber\\
\tilde{F}_{3, \, 4, \, 5, \, 6}^{Born}(\nu ,Q^{2}) \,&=&\, 0 , 	
\end{eqnarray}
where $G_M (F_2)$ are the proton magnetic (Pauli) form factors 
respectively.    
The invariant amplitude $\tilde F_2$ can be traded for $\tilde G_E$,
defined as~:
\begin{equation}
\label{Eq:Obs.11}
\tilde{G}_{E} \equiv \tilde{G}_{M}-(1+\tau )\tilde{F}_{2},
\end{equation}
which has the property that in the Born approximation it reduces to the
electric form factor, i.e. 
\begin{equation}
\tilde{G}^{Born}_{E}(\nu, Q^2)=G_{E}(Q^2).
\end{equation}
To separate the one- and two-photon exchange contributions,  
it is then useful to introduce the decompositions~: 
\begin{eqnarray}
\tilde G_M \,&=&\, G_M + \delta \tilde G_M , \\ 
\tilde G_E \,&=&\, G_E + \delta \tilde G_E. 
\end{eqnarray}
Since the amplitudes $\delta \tilde{G}_{M}$, $\delta \tilde G_E $, 
$\tilde F_{3}$, $\tilde F_{4}$, $\tilde F_{5}$, and $\tilde F_{6}$ 
vanish in Born approximation, they must   
originate from processes involving at least the exchange of two photons.  
Relative to the factor \( e^{2} \) introduced in 
Eqs.~(\ref{eq:non-flip}, \ref{eq:flip}), 
we see that they are of order \( e^{2}. \)

\section{Single spin asymmetries in elastic electron-nucleon scattering}
\label{sec:ssa}

An observable which is directly proportional to 
the two- (or multi-) photon exchange is given by the elastic scattering of an
unpolarized electron on a proton target polarized {\it normal} to the
scattering plane (or the recoil polarization normal to the
scattering plane, which is exactly the same assuming 
time-reversal invariance). 
For a target polarized perpendicular to the scattering plane, the
corresponding single spin asymmetry, which we refer to as the target normal
spin asymmetry ($A_n$), is defined by~:
\begin{eqnarray}
A_n \,=\, 
\frac{\sigma_\uparrow-\sigma_\downarrow}{\sigma_\uparrow+\sigma_\downarrow}\,,
\label{eq:tasymm}
\end{eqnarray} 
where $\sigma_\uparrow$ ($\sigma_\downarrow$) denotes the cross section 
for an unpolarized beam and for a nucleon spin 
parallel (anti-parallel) to the normal polarization vector, defined
as~:
\begin{eqnarray}
S_n^\mu \,=\, (\,0\,,\, \vec S_n \,), \hspace{2cm}
\vec S_n \,\equiv \,  (\vec{k}\times\vec{k}') \,/\, | \vec{k}\times\vec{k}' | .
\label{eq:sn}
\end{eqnarray}
\indent
As has been shown by de Rujula {\it et al.} \cite{RKR71}, 
the target (or recoil) normal spin asymmetry 
is related to the absorptive part of the elastic $e N$ scattering
amplitude (see Section~\ref{sec:impart}). 
Since the one-photon exchange amplitude 
is purely real, the leading contribution to $A_n$ is of order
$O(e^2)$, and is due to an interference between one- and two-photon
exchange amplitudes. 
\newline
\indent
When neglecting terms which correspond with electron helicity flip 
(i.e. setting $m_e = 0$), 
$A_n$ can be expressed in terms of the 
invariants for electron-nucleon elastic scattering, 
defined in Eqs.~(\ref{eq:non-flip}, \ref{eq:flip}), as~\cite{YCC04}~:
\begin{eqnarray}
A_n &=& \sqrt{\frac{2 \, \varepsilon \, (1+\varepsilon )}{\tau}} \,\,
\left( G_M^2 \,+\, \frac{\varepsilon}{\tau} \, G_E^2 \right)^{-1} \nonumber\\
&\times& \left\{ - \, G_M \, {\cal I} 
\left(\delta \tilde G_E + \frac{\nu}{M^2} \tilde F_3 \right) 
\,+ \, G_E \, {\cal I} \left(\delta \tilde G_M 
+ \left( \frac{2 \varepsilon}{1 + \varepsilon} \right) 
\frac{\nu}{M^2} \tilde F_3 \right) \right\}  \nonumber \\
&+& {\mathcal{O}}(e^4) , 
\label{eq:tnsa}
\end{eqnarray}
where $\cal I$ denotes the imaginary part. 
\newline
\indent
For a beam polarized perpendicular to the scattering plane, we can
also define a single spin asymmetry, 
analogously as in Eq.~(\ref{eq:tasymm}), where now 
$\sigma_\uparrow$ ($\sigma_\downarrow$) denotes the
cross section for an unpolarized target and for an electron beam spin 
parallel (anti-parallel) to the normal polarization vector, given by 
Eq.~(\ref{eq:sn}). 
We refer to this asymmetry as the beam normal
spin asymmetry ($B_n$). It explicitly vanishes when $m_e = 0$ as it
involves an electron helicity flip. Using the general electron-nucleon
scattering amplitude of Eqs.~(\ref{eq:non-flip},\ref{eq:flip}), 
$B_n$ is given by~\cite{GGV04}~:
\begin{eqnarray}
B_n \,&=&\, \frac{2 \, m_e}{Q} \, 
\sqrt{2 \, \varepsilon \, (1-\varepsilon )} \, \sqrt{1 + \frac{1}{\tau}} \, 
\left( G_M^2 \,+\, \frac{\varepsilon}{\tau} \, G_E^2 \right)^{-1} \nonumber\\
&\times& \left\{ - \tau \, G_M \, 
{\mathcal I}\left( \tilde F_3 
+ \frac{1}{1 + \tau} \,\frac{\nu}{M^2} \, \tilde F_5 \right)  
\, - \, G_E \, {\mathcal I} \left( \tilde{F}_{4}  
+ \frac{1}{1 + \tau} \, \frac{\nu}{M^2} \, \tilde F_5 \right) 
\right\}  \nonumber \\
\,&+& \, {\mathcal{O}}(e^4) , 
\label{eq:bnsa} 
\end{eqnarray}
As for $A_n$, we immediately see that $B_n$ vanishes in the Born approximation,
and is therefore of order $e^2$.

\section{Imaginary (absorptive) part of the 
two-photon exchange amplitude}
\label{sec:impart}

\begin{figure}[h]
\includegraphics[width=8cm]{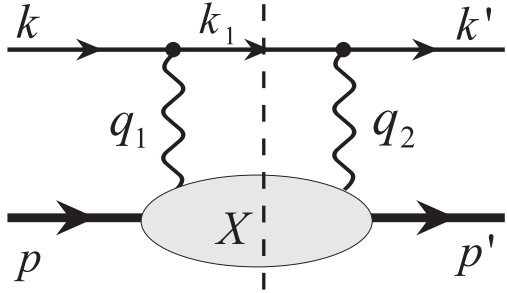}
\caption{The two-photon exchange diagram. 
The filled blob represents the response
of the nucleon to the scattering of the virtual photon.}
\label{fig:2gamma}
\end{figure}

In this section we relate the imaginary part of the
two-photon exchange amplitude to the absorptive part 
of the doubly virtual Compton scattering tensor on the nucleon, 
as shown in Fig.~\ref{fig:2gamma}. 
In the following we consider the helicity amplitudes for the 
elastic electron-nucleon scattering, 
defined in the $e^- N$ $c.m.$ frame, which are denoted by  
$T(h' , \lambda_N' \,;\, h ,\lambda_N)$. Here  
$h (h')$ denote the helicities of the initial (final) electrons
and $\lambda_N (\lambda_N')$ denote the helicities of the 
initial (final) nucleons.
These helicity amplitudes can be expressed in terms of the invariant 
amplitudes introduced in Eqs.~(\ref{eq:non-flip},\ref{eq:flip}), and the
corresponding relations can be found in Appendix~\ref{app:hel}.
These relations allow us to calculate the invariant amplitudes, once we have
constructed a model for the helicity amplitudes. 

\begin{figure}[h]
\includegraphics[width=11cm]{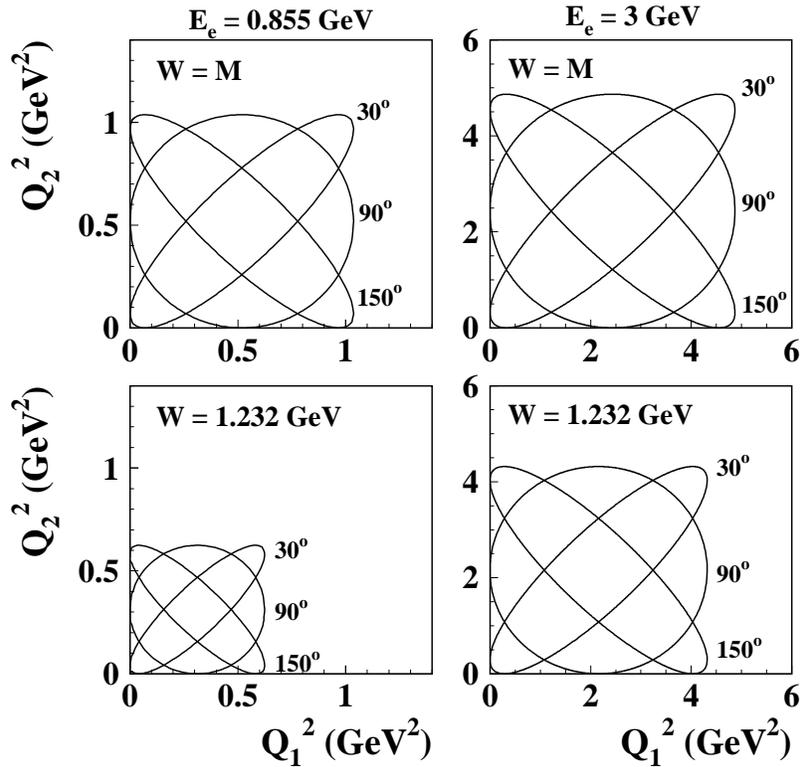}
\caption{Kinematical accessible region for the virtualities
$Q_1^2, Q_2^2$ in the phase space integral of 
Eqs.~(\ref{eq:abs}) and (\ref{eq:an2}), for  
MAMI (left panels) and JLab (right panels) kinematics for different 
{\it c.m.} angles $\theta_{c.m.}$ as indicated on the ellipses. 
The accessible regions correspond with the inside of the ellipses and 
are shown both for the elastic (upper panels) 
and for inelastic (lower panels) intermediate hadronic states. 
The intersection with the axes correspond with quasi-VCS, whereas  
the situation at $W = \sqrt{s} - m_e$ where all ellipses shrink to the point 
$Q_1^2 = Q_2^2 \simeq 0$ corresponds with quasi-RCS.}
\label{fig:qsqr_bounds}
\end{figure}

\indent
We start by calculating the discontinuity of the 
two-photon exchange amplitude, shown in Fig.~\ref{fig:2gamma}, which 
is given by
\begin{equation}
{\rm Abs} \, T_{2\gamma}\;=\;e^4\int\frac{d^3\vec{k}_1}{(2\pi)^32E_{k_1}}
\bar{u}(k',h')\gamma_\mu(\gamma \cdot k_1+m_e)\gamma_\nu u(k,h)
\frac{1}{Q_1^2Q_2^2}\cdot W^{\mu\nu}(p',\lambda_N';p,\lambda_N) \,,
\label{eq:abs}
\end{equation}
\noindent
where the momenta are defined as indicated on Fig.~\ref{fig:2gamma}, 
with $q_1 \equiv k - k_1$, $q_2 \equiv k' - k_1$, and $q_1 - q_2 = q$.
Denoting the {\it c.m.} angle between 
initial and final electrons as $\theta_{c.m.}$,
the momentum transfer $Q^2 \equiv - q^2$ in the elastic scattering 
process can be expressed as~:
\begin{eqnarray}
Q^2 = \frac{(s - M^2)^2}{2 \, s} \, \left( 1 - \cos \theta_{c.m.} \right) 
\,+\, {\mathcal{O}}(m_e^2),
\end{eqnarray}
with $s = (k + p)^2$.  
Furthermore, $Q_1^2 \equiv -q_1^2 = - (k - k_1)^2$ and 
$Q_2^2 \equiv -q_2^2 = - (k' - k_1)^2$ correspond with the 
virtualities of the two spacelike photons. 
\newline
\indent
In Eq.~(\ref{eq:abs}), the 
hadronic tensor $W^{\mu\nu}(p',\lambda_N';p,\lambda_N)$ 
corresponds with the absorptive part
of the doubly virtual Compton scattering tensor 
with two {\it space-like} photons~:
\begin{equation}
W^{\mu\nu}(p',\lambda_N';p,\lambda_N)\;=\;
\sum_X \,(2\pi)^4 \, \delta^4(p+q_1-p_X)
<p'\, \lambda_N' |J^{\dagger \mu}(0)|X> \, <X|J^\nu(0)|p \, \lambda_N>\,,
\label{eq:wtensor}
\end{equation}
\noindent
where the sum goes over all possible {\it on-shell} intermediate hadronic 
states $X$. 
Note that in the limit $p' = p$, Eq.~(\ref{eq:wtensor}) 
reduces to the forward tensor for inclusive electron-nucleon scattering 
and can be parametrized by the usual 4 nucleon forward structure 
functions. In the non-forward case however, 
the absorptive part of the 
doubly virtual Compton scattering tensor of Eq.~(\ref{eq:wtensor}) 
which enters in the evaluation of target and beam normal spin asymmetries, 
depends upon 18 invariant amplitudes~\cite{Tar75}. 
Though this may seem as a forbiddingly large number of new functions, 
we may use the unitarity relation to express the full non-forward tensor 
in terms of electroproduction amplitudes $\gamma^* N \to X$. 
The number of intermediate states $X$ which one considers in the 
calculation will then put a limit on how high in energy one can 
reliably calculate the hadronic tensor Eq.~(\ref{eq:wtensor}). 
In the following section, we will model the tensor $W^{\mu\nu}$ for the
elastic contribution ($X = N$), and in the resonance region as 
a sum over all $\pi N$ intermediate states (i.e. $X = \pi N$), 
using a phenomenological state-of-the-art calculation for the 
$\gamma^* N \to \pi N$ amplitudes.  
\newline
\indent
The phase space integral in Eq.~(\ref{eq:abs}) runs over 
the 3-momentum of the intermediate (on-shell) electron. 
Evaluating the process in the $e^- N$ {\it c.m.} system, 
we can express the {\it c.m.} momentum of the intermediate electron as~:
\begin{eqnarray}
|\vec{k}_1|^2 \,&=&\, \frac{(s - W^2 + m_e^2)^2 \,-\, 4 s m_e^2}{4 s} 
\nonumber \\
&\simeq& \frac{(s - W^2)^2}{4 s} 
\left\{1 - 2 \, m_e^2 \, \frac{(s + W^2)}{(s - W^2)^2} \right\} 
\,+\, {\mathcal{O}}(m_e^4),
\label{eq:kink1}
\end{eqnarray}
where $W^2 \equiv p_X^2$ is 
the squared invariant mass of the intermediate state
$X$. 
The {\it c.m.} momentum $|\vec k|$ of the initial (and final) electrons is 
given by the analogous expression as Eq.~(\ref{eq:kink1}) by replacing 
$W^2 \to M^2$. 
The three-dimensional phase space integral in Eq.~(\ref{eq:abs}) 
depends, besides the magnitude $|\vec k_1|$,  
upon the solid angle of the intermediate electron. 
We define the polar {\it c.m.} angle $\theta_1$ of the intermediate electron 
w.r.t. to the direction of the initial electron. The azimuthal angle 
$\phi_1$ is chosen such that $\phi_1 = 0$ 
corresponds with the scattering plane of the $e N \to e N$ process. 
Having defined the kinematics of the intermediate electron, we can 
express the virtuality of both exchanged photons. The virtuality of 
the photon with four-momentum $q_1$ is given by~:
\begin{eqnarray} 
\label{eq:q1virt} 
Q_1^2 &&\simeq \frac{1}{2 \,s} \, 
\left\{ (s - M^2) \, (s - W^2) \, (1 - \cos \theta_1) \right.  \\
&&\left. 
- m_e^2 (s + W^2) \left(1 - \frac{(s - M^2)}{(s - W^2)} \cos \theta_1 \right) 
- m_e^2 (s + M^2) \left(1 - \frac{(s - W^2)}{(s - M^2)} \cos \theta_1 \right) 
\right\} \,+\, {\mathcal{O}}(m_e^4) . \nonumber 
\end{eqnarray}
The virtuality $Q_2^2$ of the second photon has an analogous expression 
as Eq.~(\ref{eq:q1virt}) by replacing $\cos \theta_1$ by $\cos \theta_2$, 
where $\theta_2$ is the angle between the intermediate and final electrons. 
In terms of the polar and azimuthal angles $\theta_1$ and $\phi_1$ of the 
intermediate electron, one can express~:
\begin{eqnarray}
\cos \theta_2 \,=\, \sin \theta_{c.m.} \, \sin \theta_1 \, \cos \phi_1 
\,+\, \cos \theta_{c.m.} \, \cos \theta_1 .
\end{eqnarray}
\indent
In case the intermediate electron 
is collinear with the initial electron (i.e. for 
$\theta_1 \to 0$, $\phi_1 \to 0$), one obtains from Eq.~(\ref{eq:q1virt}) that 
both photon virtualities are given by~:
\begin{eqnarray}
Q_{1, \, VCS}^2 \,&\equiv& \, Q_1^2 (\theta_1 = 0, \phi_1 = 0) 
\,\simeq\, m_e^2 \, \frac{(W^2 - M^2)^2}{(s - W^2)\,(s - M^2)}, \nonumber \\ 
Q_{2, \, VCS}^2 \,&\equiv& \, Q_2^2 (\theta_1 = 0, \phi_1 = 0) 
\,\simeq\, \frac{(s - W^2)}{(s - M^2)}\, Q^2 \,+\, {\mathcal{O}}(m_e^2). 
\end{eqnarray}
Note that when the intermediate and initial electrons are collinear, 
then also the photon with momentum $\vec q_1 = \vec k - \vec k_1$ is 
collinear with this direction. 
For the elastic case ($W = M$) this precisely corresponds with 
the situation where the first photon is soft (i.e. $q_1 \to 0$) and 
where the second photon carries the full momentum transfer 
$Q_2^2 \simeq Q^2$.  
For the inelastic case ($W > M$) the first photon is 
hard but becomes quasi-real (i.e. $Q_1^2 \sim m_e^2$). 
In this case, the virtuality of the second photon is smaller than $Q^2$. 
An analogous situation occurs when the intermediate electron is 
collinear with the final electron 
(i.e. $\theta_2 \to 0$, $\phi_1 \to 0$, which is equivalent with 
$\theta_1 \to \theta_{c.m.}$). 
These kinematical situations with one quasi-real photon and one virtual photon 
correspond with quasi virtual Compton scattering (quasi-VCS), 
and correspond at the lepton side with the Bethe-Heitler process, see  
e.g. Ref.~\cite{GV98} for details.
\newline
\indent
Besides the near singularities corresponding with quasi-VCS, where  
the intermediate electron 
is collinear with either the incoming or outgoing electrons, 
the two photon exchange process also has a near singularity when 
the intermediate electron momentum goes to zero $|\vec k_1| \to 0$ 
(i.e. the intermediate electron is soft). 
In this case the first photon takes on the full momentum of the 
initial electron, i.e. $\vec q_1 \to \vec k$, whereas the 
second photon takes on the full momentum of the final electron, 
i.e. $\vec q_2 \to \vec k'$.  
One immediately sees from Eq.~(\ref{eq:kink1}) 
that this situation occurs when the invariant mass of the hadronic 
state takes on its maximal value $W_{max} = \sqrt{s} - m_e$.   
In this case, both photon virtualities are given by~:
\begin{eqnarray}
Q_{1, \, RCS}^2 \,=\, Q_{2, \, RCS}^2 \,
\simeq\, m_e \, \frac{(s - M^2)}{\sqrt{s}} \,\left(1 - \cos \theta_1 \right). 
\label{eq:q1real}
\end{eqnarray}
This kinematical situation with two quasi-real photons corresponds with 
quasi-real Compton scattering (quasi-RCS).
\newline
\indent 
Due to the near singularities in the phase space integral of 
Eq.~(\ref{eq:abs}) corresponding with the quasi-VCS and quasi-RCS processes, 
special care was taken when integrating over these regions, as the 
integrand varies strongly over regions governed by the electron mass.  
Below we will show that these near singularities 
may give important contributions (logarithmic enhancements) 
under some kinematical conditions. 
In Fig.~\ref{fig:qsqr_bounds}, we show the full 
kinematical accessible region for the virtualities
$Q_1^2, Q_2^2$ in the phase space integral of Eq.~(\ref{eq:abs}).
\newline
\indent
The normal spin asymmetries $A_n$ and $B_n$,
 discussed in Sections~\ref{sec:ssa}, are a direct
measure of the absorptive part of the two-photon exchange amplitude 
and can be expressed as \cite{RKR71}~:
\begin{equation}
A_n\;=\;
\frac{2 \, {\rm Im}(\sum_{spins}T_{1\gamma}^*\cdot {\rm Abs} \,T_{2\gamma})}
{\sum_{spins}|T_{1\gamma}|^2}\, , 
\label{eq:an1}
\end{equation}
where $T_{1\gamma}$ denotes the one-photon exchange amplitude. 
Using Eq.~(\ref{eq:abs}), we can express Eq.~(\ref{eq:an1}) in terms of a
3-dimensional phase-space integral~:
\beqn
A_n \;=\; -\frac{1}{(2\pi)^3}\frac{e^2Q^2}{D(s,Q^2)}
\int_{M^2}^{(\sqrt{s}-m_e)^2}dW^2\,\frac{|\vec{k}_1|}{4 \, \sqrt{s} }
\int d\Omega_{k_1}\frac{1}{Q_1^2 \, Q_2^2}\,{\rm Im}
\left\{L_{\alpha\mu\nu} \, H^{\alpha\mu\nu}\right\}\,.
\label{eq:an2}
\eeqn
The denominator factor $D(s, Q^2)$ in Eq.~(\ref{eq:an2}) is given through the 
one-photon exchange cross section as~:  
\beqn
 D(s,Q^2) \;\equiv\; \frac{Q^4}{e^4} \cdot \sum_{spins}|T_{1\gamma}|^2 
\,=\, \, 8 \, \,\frac{(4 M^2 \tau)^2}{1 - \varepsilon} \, \, 
\left\{\, G_M^2 \,+\, \frac{\varepsilon}{\tau} \, G_E^2 \, \right\} .
\eeqn
\noindent
Furthermore in Eq.~(\ref{eq:an2}), 
the leptonic ($L_{\alpha\mu\nu}$) and hadronic ($H^{\alpha\mu\nu}$) 
tensors are given by~:
\beqn
L_{\alpha\mu\nu}&=&
\bar{u}(k',h')\gamma_\mu (\gamma \cdot k_1 + m_e) \gamma_\nu u(k,h)
\cdot
\left[\bar{u}(k',h')\gamma_\alpha u(k,h)\right]^*\, , 
\label{eq:lept} \\
H^{\alpha\mu\nu}&=& W^{\mu\nu}\cdot
\left[\bar{u}(p',\lambda_N')\Gamma^\alpha(p', p) u(p,\lambda_N)\right]^* \, ,
\label{eq:hadt}
\eeqn
where 
\begin{eqnarray}
\Gamma^\alpha(p', p) \equiv G_M \gamma^\alpha - F_2 \, P^\alpha / M. 
\label{eq:gamma}
\end{eqnarray}
In Eqs.~(\ref{eq:lept}, \ref{eq:hadt}) a sum is understood over the
helicities of the unpolarized particles. 
To evaluate the target normal spin asymmetry $A_n$, 
we need the unpolarized lepton tensor, which is given by (neglecting the 
terms proportional to the electron mass $m_e$)~:
\beqn
L_{\alpha\mu\nu}^{unpol}&=& \mathrm{Tr} 
\left\{ \sl{k'} \gamma_\mu \sl{k_1} \gamma_\nu \sl{k} \gamma_\alpha \right\}.
\label{eq:lept1}
\eeqn
The beam normal spin asymmetry $B_n$ involves 
the polarized lepton tensor, which is given by~:
\beqn
L_{\alpha\mu\nu}^{pol}&=& \mathrm{Tr} 
\left\{ (\sl{k'} + m_e) \gamma_\mu (\sl{k_1} + m_e) \gamma_\nu 
\gamma_5 \sl{\xi} ( \sl{k} + m_e) \gamma_\alpha \right\},
\label{eq:lept2}
\eeqn
where $\xi^\beta$ is the polarization vector, for an electron polarized 
normal to the scattering plane.  
We see from Eq.~(\ref{eq:lept2}) that the polarized lepton tensor 
vanishes for massless electrons. Keeping only the leading term in $m_e$, it is 
given by~:
\beqn
L_{\alpha\mu\nu}^{pol}&=& m_e \left\{ \,  
- \mathrm{Tr} 
\left\{ \gamma_5 \gamma_\mu \sl{k_1} \gamma_\nu \sl{\xi} \sl{k} 
\gamma_\alpha \right\}
+ \mathrm{Tr} 
\left\{ \gamma_5 \sl{k'} \gamma_\mu \sl{k_1} \gamma_\nu \sl{\xi}  
\gamma_\alpha \right\}
- \mathrm{Tr} 
\left\{ \gamma_5 \sl{k'} \gamma_\mu \gamma_\nu \sl{\xi} \sl{k} 
\gamma_\alpha \right\} 
\, \right\} \nonumber \\
&+&  \, {\mathcal{O}}(m_e^2) .
\label{eq:lept3}
\eeqn

\section{Models for the hadronic tensor}
\label{sec:models}

In this section, we discuss several models for the non-forward 
hadronic tensor $W^{\mu \nu}$ of Eq.~(\ref{eq:wtensor}) which 
enters in the imaginary part of the two-photon exchange amplitude.  
These models will be used further on to evaluate the target and 
beam normal spin asymmetries. 
\newline
\indent
An initial guess, is to approximate the non-forward tensor by the 
corresponding forward tensor in terms of 4 nucleon structure functions, 
as was done in the calculations of Ref.~\cite{RKR71}, and 
adapted in Ref.~\cite{AAM02} by complementing 
the nucleon structure functions by a form factor dependence.
This may be a reliable first estimate  
when one is interested in the kinematical 
limit of high energy and very small momentum transfer ($Q^2 \ll s$), 
corresponding with the Regge regime. The SLAC E158 experiment~\cite{E158}, 
which corresponds with $E_\gamma \simeq 50$~GeV and $Q^2 \leq 0.05$~GeV$^2$, 
accesses this diffractive region and 
may be a good testing ground for such models.    
\newline
\indent
To go beyond the very forward angle approximation for the hadronic tensor, 
and in order to compare quantitatively with beam normal spin asymmetry 
measurements performed or in progress 
at MIT-Bates~\cite{Wells01}, MAMI~\cite{Maas03}, and JLab~\cite{Happex,G0} 
at intermediate beam energies in the 1 GeV region, 
one immediately faces the full complexity of the non-forward 
doubly virtual Compton scattering tensor. This non-forward tensor 
can be parametrized in general in terms of 
18 invariant amplitudes~\cite{Tar75}. 
As we are interested in this work in the absorptive part of the non-forward 
doubly virtual Compton scattering tensor, 
we may use the unitarity relation to express the full non-forward tensor 
in terms of electroabsorption amplitudes $\gamma^* N \to X$ 
at different photon virtualities. 
This same strategy has been used before in the description of 
real and virtual Compton scattering in the resonance region, 
and checked against data in Ref.~\cite{DPV03}. 
We will subsequently model the non-forward tensor $W^{\mu\nu}$ for the
elastic contribution ($X = N$), and in the resonance region as 
a sum over all $\pi N$ intermediate states (i.e. $X = \pi N$).

\subsection{Elastic contribution}

The elastic contribution to $W^{\mu \nu}$,
corresponding with the nucleon  
intermediate state in the blob of Fig.~\ref{fig:2gamma},
is exactly calculable in terms of on-shell nucleon electromagnetic 
form factors as~:
\beqn
W^{\mu\nu}_{el}(p',\lambda_N'; p,\lambda_N) 
\;&=&\;2\pi \, \delta(W^2-M^2) \nonumber \\
&\times& \, \bar{u}(p',\lambda_N') \, \Gamma^\mu(p', p_X) \, 
(\gamma \cdot p_X+ M) \, \Gamma^\nu(p_X, p) \, u(p,\lambda_N)\, ,
\label{eq:welast}
\eeqn
where $\Gamma^\mu$ is given as in Eq.~(\ref{eq:gamma}).

\subsection{Inelastic contribution : sum over $\pi N$ intermediate states 
using the MAID model (resonance region)}

The inelastic contribution to $W^{\mu\nu}$ corresponding with the $\pi N$ 
intermediate states in the blob of Fig.~\ref{fig:2gamma}, 
is given by

\begin{eqnarray}
W^{\mu\nu}(p',\lambda'_N;p,\lambda_N)&=&
\frac{1}{4\pi^2}\frac{|\vec p_\pi|^2}{[|\vec p_\pi|(E_\pi+E_n)+E_\pi|\vec k_1|\hat k_1\cdot \hat p_\pi]}\nonumber\\
& &\times \sum_{\lambda_n}\int{\rm d}\Omega_\pi \; 
\bar u(p',\lambda'_N) J^{\dagger \mu}_{\pi N}
u(p_n,\lambda_n) \; 
\bar u(p_n,\lambda_n)
J^\nu_{\pi N} u(p,\lambda_N),
\label{eq:maidtensor}
\end{eqnarray}
where $p_\pi=(E_\pi,\vec p_\pi)$ and $p_n=(E_n,\vec{p}\,_n)$
 are the four-momenta of the intermediate pion and nucleon states 
respectively, and $\vec k_1 = - \vec p_\pi - \vec p_n$. 
In Eq.~(\ref{eq:maidtensor}) the integration runs over the polar and azimuthal 
angles of the intermediate pion, 
and $J^\nu_{\pi N}$  and $J^{\dagger\mu}_{\pi N}$ are the 
pion electroproduction currents,  describing the 
excitation and de-excitation of the $\pi N$ intermediate state, respectively.
Following Ref.~\cite{Ber67}, we parametrize the 
matrix element of the pion electroproduction current 
in terms of six invariant amplitudes $A_i$ as~:
\begin{eqnarray}
\bar u(p_n,\lambda_n)\,J^\nu_{\pi N} \, u(p,\lambda_N) = 8 \pi W \,\, 
\bar u(p_n,\lambda_n)
 \sum_{i = 1}^6 A_i(W^2,t_\pi,Q_1^2)\, M^\nu_i\, u(p,\lambda_N),
\label{eq:current}
\end{eqnarray}
where $W^2=(p+q_1)^2=(p_\pi+p_n)^2$ is the squared {\it c.m.} energy of the 
$\pi N$ system,  
$t_\pi=(p-p_n)^2=(p_\pi-q_1)^2$ is the squared four-momentum transfer 
in the $\gamma N \to \pi N$ process. 
In Eq.~(\ref{eq:current}), the covariants $M^\nu_i$ are given by 
\begin{eqnarray}
M^\nu_1&=&
-\frac{1}{2}i\gamma_5\left(\gamma^\nu\sl{q}_1-\sl{q}_1\gamma^\nu\right)\, ,
\nonumber\\
M^\nu_2&=&2i\gamma_5\left(P_{in}^\nu\, q_1\cdot(p_\pi-\frac{1}{2}q_1)-
(p_\pi-\frac{1}{2}q_1)^\nu q_1\cdot P_{in}\right)\, ,\nonumber\\
M^\nu_3&=&-i\gamma_5\left(\gamma^\nu\, q_1\cdot p_\pi
-\sl{q}_1p_\pi^\nu\right)\, ,\nonumber\\\
M^\nu_4&=&-2i\gamma_5\left(\gamma^\nu\, q_1\cdot P_{in}
-\sl{q}_1P_{in}^\nu\right)-2 M \, M^\nu_1\, ,\nonumber\\\
M^\nu_5&=&i\gamma_5\left(q_1^\nu\, q_1\cdot p_\pi
+Q_1^2 p_\pi^\nu\right)\, ,\nonumber\\\
M^\nu_6&=&-i\gamma_5\left(\sl{q}_1q_1^\nu+Q_1^2
\gamma^\nu\right)\, ,
\end{eqnarray}
where $P_{in}^\nu= (p+p_n)^\nu / 2$, and $\sl{a}=a_\nu\gamma^\nu$. 
Analogous expressions hold for the pion electroproduction current for the 
second virtual photon. 
\newline
\indent
For the calculation of the invariant amplitudes $A_i,$ we use the 
phenomenological MAID analysis (version 2000)~\cite{Tia01}, 
which contains both resonant 
and non-resonant pion production mechanisms.

\section{Results and discussion}
\label{sec:results}

\begin{figure}[h]
\includegraphics[width=14cm]{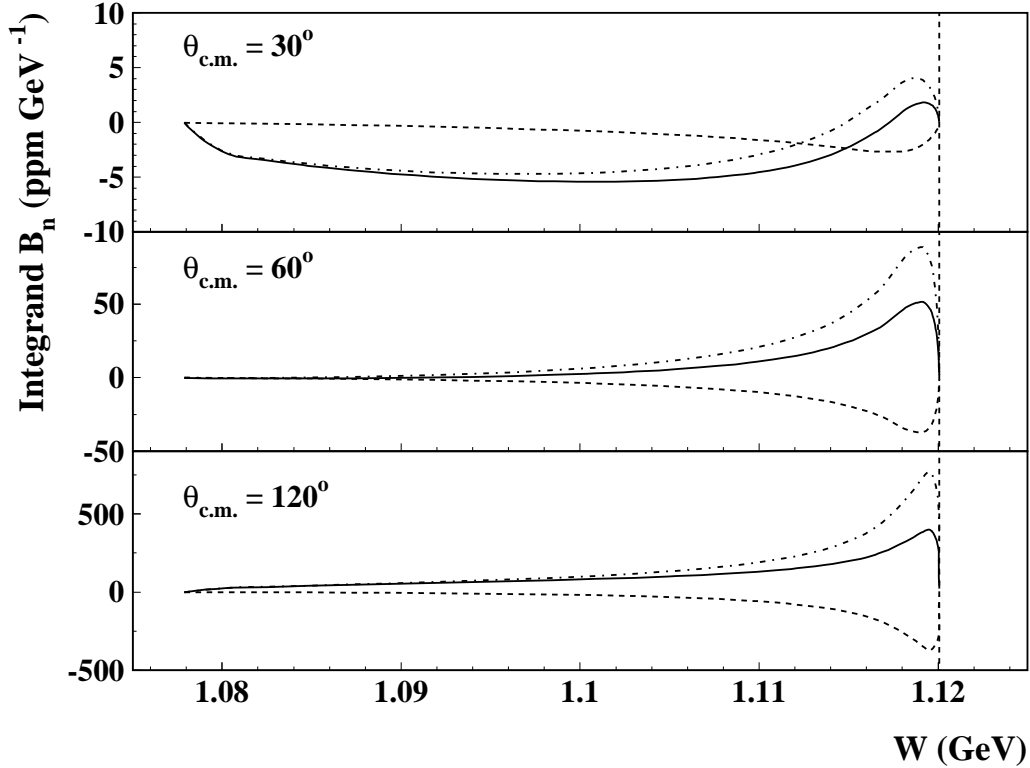}
\vspace{-1cm}
\caption{
Integrand in $W$ of the beam normal spin asymmetry $B_n$ 
for $e^{- \uparrow} p \to e^- p$ 
at a beam energy of $E_e=0.2$ GeV and at different $c.m.$ scattering angles 
as indicated on the figure.
The dashed curves are the contribution from the $\pi^0 p$ channel,
the dashed-dotted curves show the contribution from the $\pi^+ n$ 
channel, and the solid curves are the sum of the contributions from the 
$\pi^+ n$ and $\pi^0 p$ channels. 
The vertical dashed line indicates the upper limit of the $W$ integration, 
i.e. $W_{max} = \sqrt{s} - m_e$.}
\label{fig:int_bn1}
\end{figure}

\begin{figure}[h]
\includegraphics[width=14cm]{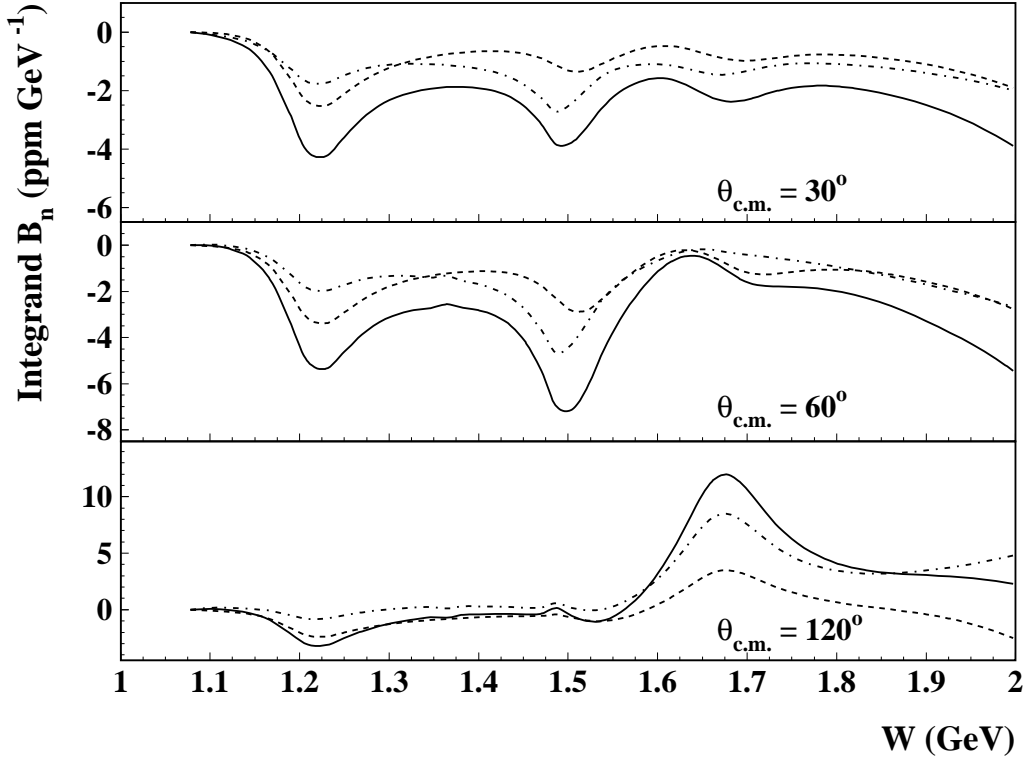}
\vspace{-1cm}
\caption{
Integrand in $W$ of the beam normal spin asymmetry $B_n$ 
for $e^{- \uparrow} p \to e^- p$ 
at a beam energy of $E_e=3$ GeV and at different $c.m.$ scattering angles 
as indicated on the figure. 
The dashed curves are the contribution from the $\pi^0 p$ channel,
the dashed-dotted curves show the contribution from the $\pi^+ n$ 
channel, and the full curves are the sum of the contributions from the 
$\pi^+ n$ and $\pi^0 p$ channels.
Note that for $E_e = 3$~GeV, the upper integration range in $W$ 
is given by $W_{max} \simeq 2.55$~GeV.}
\label{fig:int_bn4}
\end{figure}

\begin{figure}[h]
\includegraphics[width=14cm]{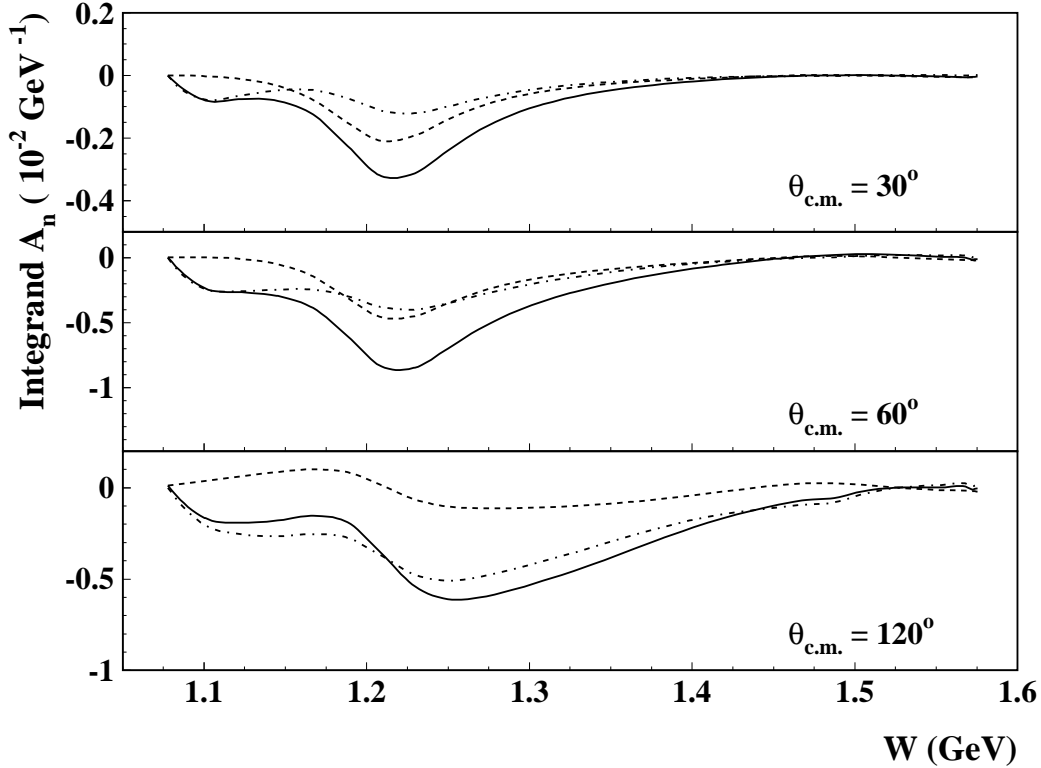}
\vspace{-1cm}
\caption{
Integrand in $W$ of the target normal spin asymmetry $A_n$ 
for $e^- p^\uparrow \to e^- p$
for a beam energy of $E_e=0.855$ GeV and at different $c.m.$ scattering 
angles as indicated on the figure. 
The dashed curves are the contribution from the $\pi^0 p$ channel,
the dashed-dotted curves show the contribution from the $\pi^+ n$ 
channel, and the solid curves are the sum of the  
$\pi^+ n$ and $\pi^0 p$ channels.}
\label{fig:int_an1}
\end{figure}

In this section we show our results for both beam and target normal spin 
asymmetries for elastic electron-proton and electron-neutron scattering. 
We estimate the non-forward hadronic tensor entering the 
two-photon exchange amplitude through nucleon (elastic contribution) 
and $\pi N$ intermediate states (inelastic contribution) as described above. 
Our calculation covers the whole resonance region, using phenomenological 
$\pi N$ electroproduction amplitudes as input, and addresses  
measurements performed or in progress 
at MIT-Bates~\cite{Wells01}, MAMI~\cite{Maas03} and JLab~\cite{Happex,G0}, 
where the beam energies are below 1 GeV or in the 1-3 GeV range.
\begin{figure}[h]
\includegraphics[width=11cm]{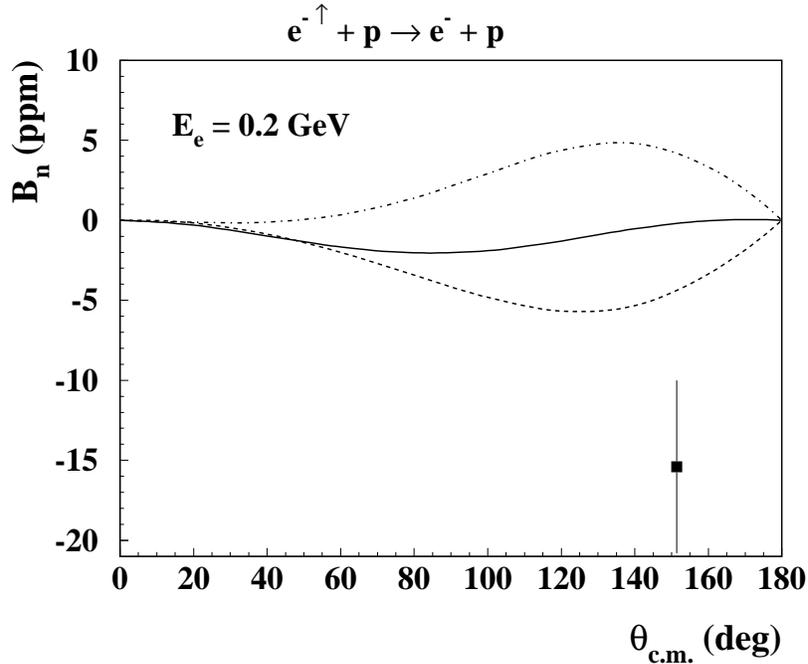}
\caption{Beam normal spin asymmetry $B_n$ for $e^{- \uparrow} p \to e^- p$ 
at a beam energy $E_e = 0.2$~GeV 
as function of the $c.m.$ scattering angle,  
for different hadronic intermediate states ($X$) in the blob  of
Fig.~\ref{fig:2gamma} ~: 
$N$ (dashed curve),  
$\pi N$  (dashed-dotted curve), 
sum of the $N$ and $\pi N$ (solid curve). 
The data point is from the SAMPLE Collaboration (MIT-Bates)~\cite{Wells01}.}
\label{fig:bn_mit}
\end{figure}
\newline
\indent
In Fig.~\ref{fig:bn_mit}, we show the beam normal spin asymmetry $B_n$ 
for elastic $e^{- \uparrow} p \to e^- p$ scattering at a low beam energy of 
$E_e = 0.2$~GeV. At this energy, the elastic contribution (where the 
hadronic intermediate state is a nucleon) is sizeable. The inelastic 
contribution is dominated by the region of threshold pion production, 
as is shown in Fig.~\ref{fig:int_bn1}, where we display the integrand 
of the $W$-integration for $B_n$. When integrating the full curve in 
Fig.~\ref{fig:int_bn1} over $W$, one obtains the total inelastic contribution 
to $B_n$ (i.e. dashed-dotted curve in Fig.~\ref{fig:bn_mit}).  
One sees from Fig.~\ref{fig:int_bn1} that at backward {\it c.m.} angles 
(i.e. with increasing $Q^2$) the $\pi^+ n$ and $\pi^0 p$ intermediate 
states contribute with opposite sign. 
Such a behavior is because the non-forward hadronic tensor involves 
electroproduction amplitudes at different virtualities. 
It would be absent when approaching the non-forward tensor by a  
forward tensor in terms of unpolarized structure functions, 
because the positivity of the unpolarized structure functions requires 
all channels to contribute with the same sign.  
Furthermore, one notices in Fig.~\ref{fig:int_bn1} that 
the peaked structure at the maximum 
possible value of the integration range in $W$, 
i.e. $W_{max} = \sqrt{s} - m_e$, is due to the near singularity 
(in the electron mass) 
corresponding with quasi-RCS as discussed in Section~\ref{sec:impart}. 
\begin{figure}[h]
\includegraphics[width=12cm]{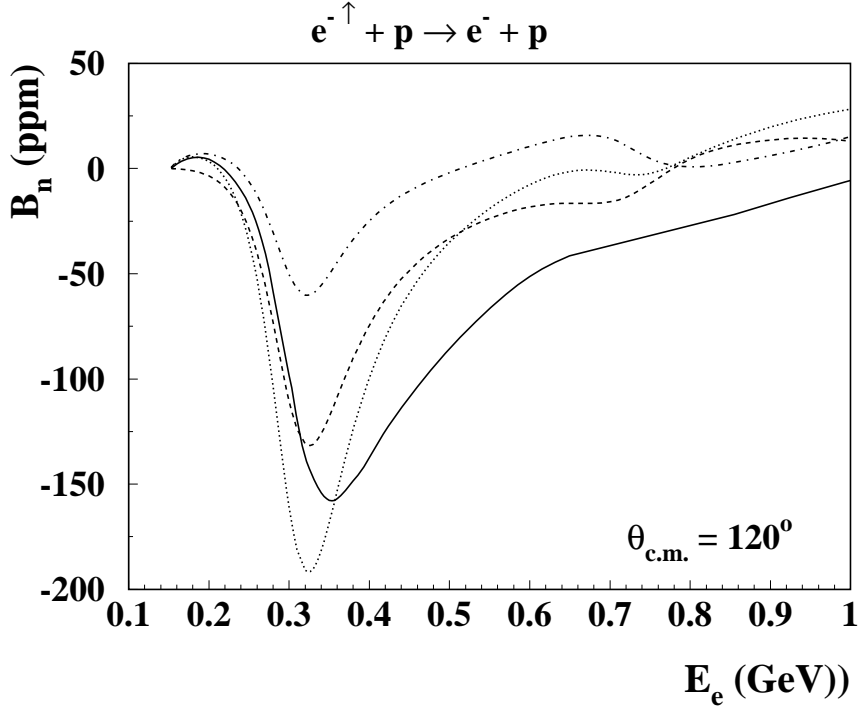}
\caption{
Beam energy dependence of the beam normal spin asymmetry $B_n$ 
for $e^{- \uparrow} p \to e^- p$ 
at fixed scattering angle $\theta_{c.m.} = 120^o$. 
The solid curve is the total inelastic calculation for both 
$\pi^+ n$ and $\pi^0 p$ channels. 
The other three curves are obtained by approximating the hadronic tensor 
$W^{\mu \nu}$  in Eqs.~(\ref{eq:an2},\ref{eq:hadt}) by its value 
at $W = W_{max} = \sqrt{s} - m_e$, corresponding with the 
quasi-real Compton scattering contribution 
for the $\pi^0 p$ channel (dashed curve),
for the $\pi^+ n$ channel (dashed-dotted curve),
and for the sum of $\pi^0 p$ and $\pi^+ n$ channels (dotted curve).}
\label{fig:int_endep}
\end{figure}
\newline
\indent
We investigate the contribution of this quasi-real Compton scattering 
to the total asymmetry $B_n$ as function of the beam energy 
at a backward angle $\theta_{cm} = 120^o$ in Fig.~\ref{fig:int_endep}. 
In this figure, we compare the full calculation (solid curve) with an 
approximate calculation (dotted curve) where the hadronic tensor 
$W^{\mu \nu}$ in Eq.~(\ref{eq:an2}, \ref{eq:hadt}) is evaluated 
at the end-point $W_{max}$, and can subsequently 
be taken out of the $W$-integration. 
This calculation corresponds with the quasi-RCS 
contribution to $B_n$. It is seen from 
Fig.~\ref{fig:int_endep} that for energies up to about $E_e \simeq 0.4$~GeV, 
the quasi-real Compton scattering is dominating the total result. It is 
also seen that when approaching the $\pi N$ threshold there is a sign change 
in $B_n$ which is driven by the non-resonant $\pi^+ n$ production process 
which yields a positive integrand around threshold. The threshold region 
in the present calculation (MAID) is consistent with chiral symmetry 
predictions, and is therefore largely model independent. 
It is seen from Fig.~\ref{fig:bn_mit} that the inelastic 
and elastic contributions at a low energy of 0.2 GeV have opposite sign, 
resulting in quite a small asymmetry around this particular energy. 
It is somewhat puzzling that the only experimental data point at this energy 
indicates a larger negative value at backward angles, 
although with quite large error bar. 
\begin{figure}[h]
\includegraphics[width=14cm]{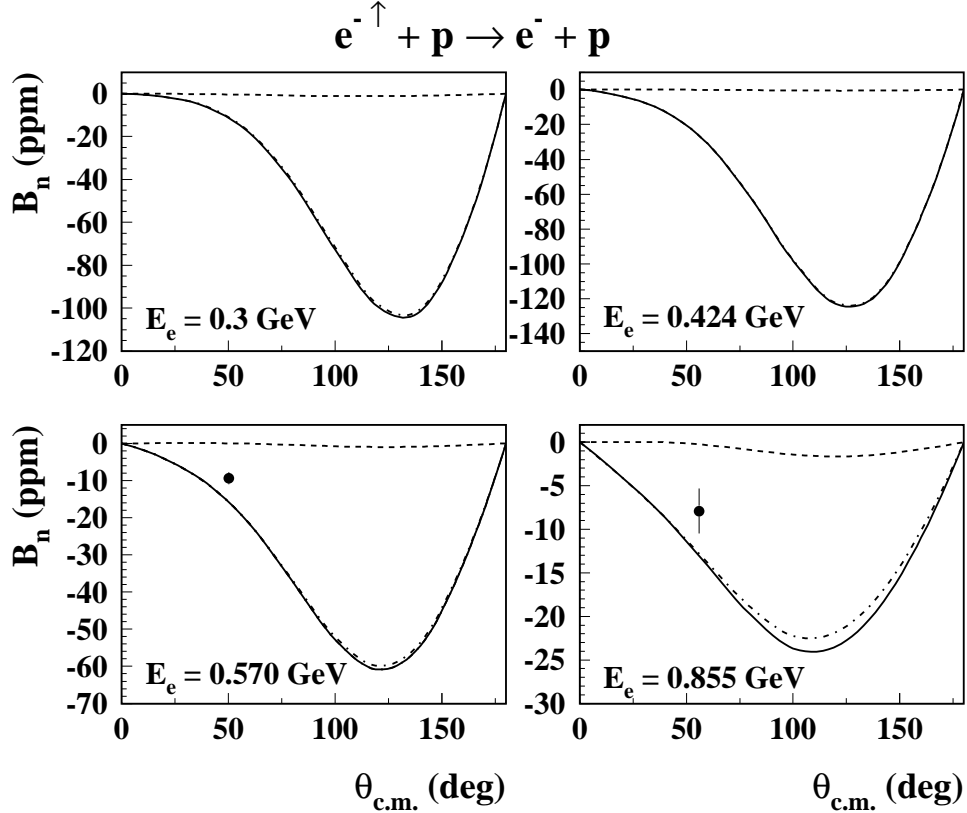}
\caption{Beam normal spin asymmetry $B_n$ 
for $e^{- \uparrow} p \to e^- p$ 
as function of the $c.m.$ scattering angle  
at different beam energies, as indicated on the figure.  
The calculations are for different hadronic intermediate 
states ($X$) in the blob of Fig.~\ref{fig:2gamma} ~: 
$N$ (dashed curve),  
$\pi N$  (dashed-dotted curves), 
sum of the $N$ and $\pi N$ (solid curves). 
The data points are from the 
A4 Collaboration (MAMI)~\cite{Maas03}.}
\label{fig:bn_mami}
\end{figure}
\newline
\indent
In Fig.~\ref{fig:bn_mami}, we show $B_n$ at different beam energies below 
$E_e = 1$~GeV. It is clearly seen that at energies $E_e = 0.3$~GeV and higher  
the elastic contribution yields only a very small relative 
contribution. Therefore $B_n$ is a direct measure of the inelastic part which 
gives rise to sizeable large asymmetries, of the order of several tens of ppm 
in the backward angular range, mainly driven by the quasi-RCS 
near singularity.  
At forward angles, the size of the predicted  
asymmetries is compatible with the first 
high precision measurements performed at MAMI. It will be worthwhile to 
investigate if the slight overprediction (in absolute value) of $B_n$, 
in particular at $E_e = 0.57$~GeV, is also seen in a backward angle 
measurement, which is planned in the near future at MAMI. 
\begin{figure}[h]
\includegraphics[width=14cm]{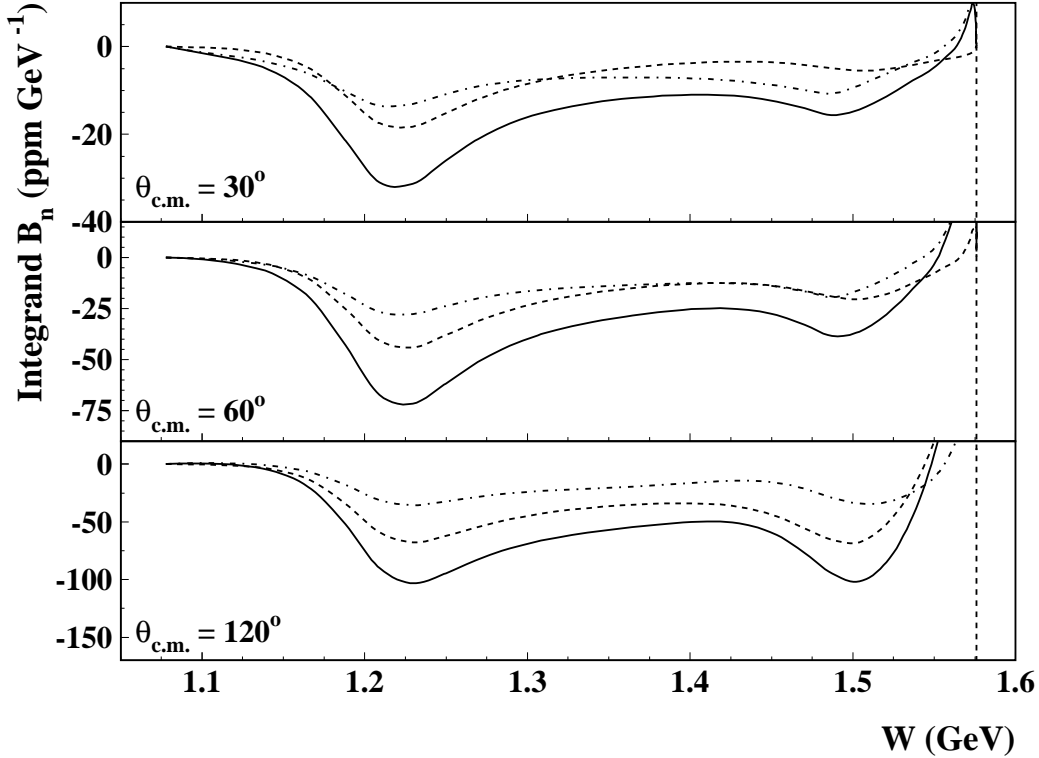}
\vspace{-1cm}
\caption{
Integrand in $W$ of the beam normal spin asymmetry $B_n$ 
for $e^{- \uparrow} p \to e^- p$ 
at a beam energy of $E_e=0.855$ GeV and at different $c.m.$ 
scattering angles as indicated on the figure. 
The dashed curves are the contribution from the $\pi^0 p$ channel,
the dashed-dotted curves show the contribution from the $\pi^+ n$ 
channel, and the solid curves are the sum of the contributions from the 
$\pi^+ n$ and $\pi^0 p$ channels.
The vertical dashed line indicates the upper limit of the $W$ integration, 
i.e. $W_{max} = \sqrt{s} - m_e$.}
\label{fig:int_bn2}
\end{figure}
\newline
\indent
To gain a better understanding of how the inelastic 
contribution to $B_n$ arises, 
we show in Fig.~\ref{fig:int_bn2} the integrand of $B_n$ 
at $E_e = 0.855$~GeV and at different scattering angles. The resonance 
structure is clearly reflected in the integrands for both 
$\pi^+ n$ and $\pi^0 p$ channels. At forward angles, the quasi-real 
Compton scattering at the endpoint $W = W_{max}$ only yields a very 
small contribution, which grows larger when going to backward angles. 
This quasi-RCS contribution is of opposite sign as the 
remainder of the integrand, and therefore determines the position of the 
maximum (absolute) value of $B_n$ when going to backward angles. 
\begin{figure}[h]
\includegraphics[width=12cm]{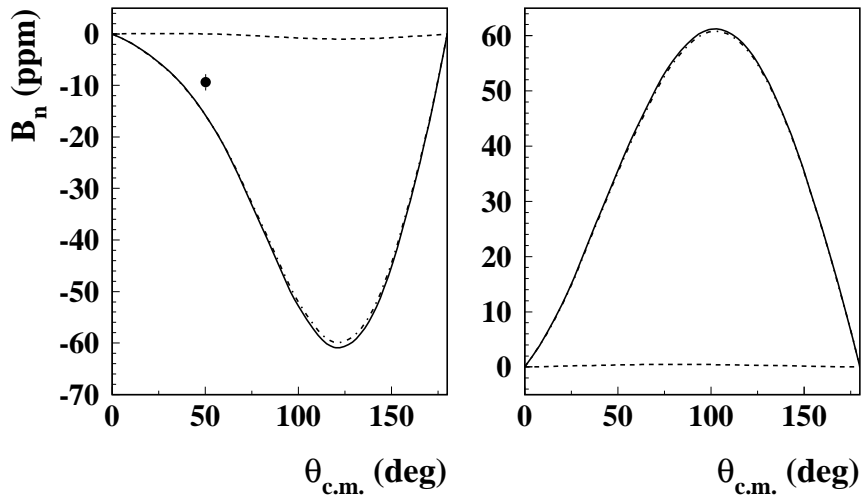}
\caption{Comparison of the beam normal spin asymmetry $B_n$ for the processes  
$e^{- \uparrow} p \to e^- p$ (left panel), and  
$e^{- \uparrow} n \to e^- n$ (right panel) 
at beam energy $E_e=0.570$ GeV
as function of the $c.m.$ scattering angle,  
for different hadronic intermediate states ($X$) in the blob  of
Fig.~\ref{fig:2gamma} ~: 
$N$ (dashed curve),  
$\pi N$  (dashed-dotted curve), 
sum of $N$ and $\pi N$ (solid curve).}
\label{fig:bnsa_neutron}
\end{figure}
\newline
\indent
In Fig.~\ref{fig:bnsa_neutron}, 
we compare the beam normal spin asymmetries at $E_e = 0.570$~GeV 
for both proton and neutron. 
It is seen that the proton and neutron values of $B_n$ are of opposite 
sign and similar in magnitude. This can be understood from 
Eq.~(\ref{eq:bnsa}) and noting that the term proportional to $G_M$ 
dominates $B_n$. As the magnetic form factor $G_M$ changes sign between 
proton and neutron, and because the two-photon 
exchange amplitudes in the $\Delta$ region (isovector transition) 
have the same sign and magnitude between proton and neutron, one obtains a 
beam normal spin asymmetry of similar magnitude and opposite sign 
between both cases.   
\begin{figure}[h]
\includegraphics[width=12cm]{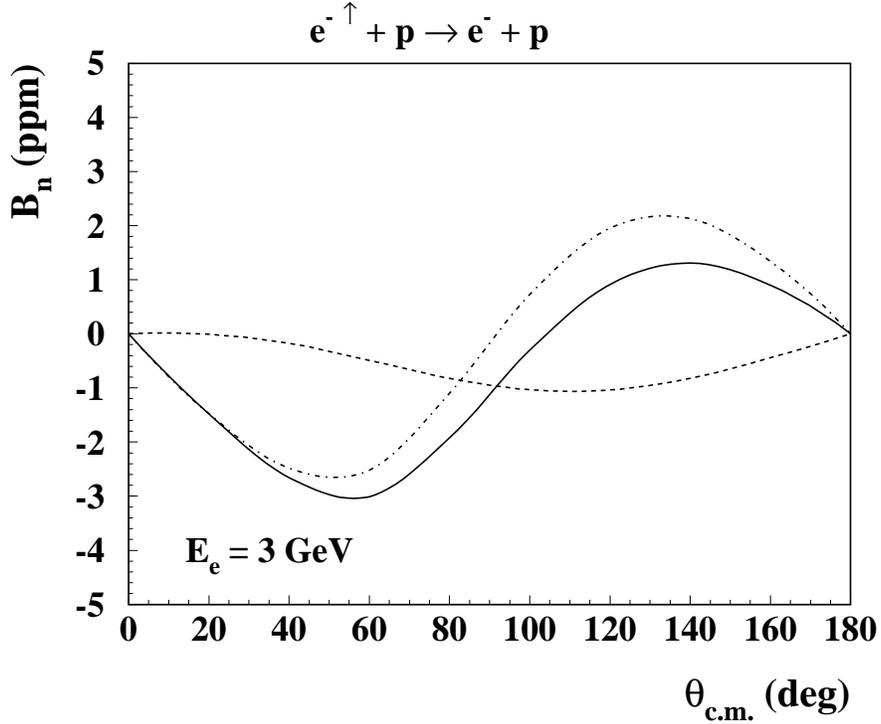}
\caption{Beam normal spin asymmetry $B_n$ 
for $e^{- \uparrow} p \to e^- p$ at a beam energy $E_e=3$ GeV
as function of the $c.m.$ scattering angle,  
for different hadronic intermediate states ($X$) in the blob  of
Fig.~\ref{fig:2gamma} ~: 
$N$ (dashed curve),  
$\pi N$  (dashed-dotted curve), 
sum of $N$ and $\pi N$ (solid curve). For the $\pi N$ intermediate states, 
we estimates the total contribution for $W \leq 2$~GeV.}
\label{fig:bn_3p0}
\end{figure}
\newline
\indent
In Fig.~\ref{fig:bn_3p0}, we show our results for the beam normal 
spin asymmetry at $E_e = 3$~GeV where parity violation programs 
are underway at JLab (G0 \cite{G0} and Happex-2 \cite{Happex} experiments). 
One notices from Fig.~\ref{fig:bn_3p0} that the elastic contribution 
at $E_e = 3$~GeV is negligibly small at forward angles, and reaches its 
largest value (in magnitude) of around -1 ppm in the backward angular range. 
The inelastic part is calculated using $\pi N$ intermediate states for 
$W < 2$~GeV. The inelastic contribution to $B_n$ displays an interesting 
structure as it is negative (around -3 ppm) 
in the forward angular range and changes sign around 
$\theta_{c.m.} \simeq 90^o$. This can be understood by comparing the 
$W$-dependence of the integrands of $B_n$ between forward and backward 
angular situations, as is shown on Fig.~\ref{fig:int_bn4}. 
The integrand of $B_n$ displays three prominent 
resonance structures corresponding with the $\Delta(1232)$ and 
dominantly with the $D_{13}(1520)$ and $F_{15}(1680)$ resonances. 
At a forward angle, all three resonance regions enter with the same sign in 
$B_n$. At a backward angle (see panel for $\theta_{c.m.} = 120^o$) however, 
one sees that the first two resonance regions are largely damped whereas 
the third resonance region shows up prominently and yields a contribution 
to $B_n$ with opposite sign. This can in turn be understood because 
at more backward angles at fixed $W$, the integration range for $B_n$ 
is dominated by the quasi-VCS regions, where one of the photons has a 
larger virtuality than at forward angle, as is seen on 
Fig.~\ref{fig:qsqr_bounds}. At larger photon virtuality, the first two 
resonance regions drop faster with $Q^2$ than the third region as follows 
from phenomenological pion electroproduction analyses and as is built into 
the MAID amplitudes. Furthermore, the sign change of the third resonance 
region at backward angles again stresses the importance to model the 
full non-forward Compton tensor. This sign change, as follows from the MAID 
model, is similar as the corresponding sign change with increasing $Q^2$ for 
the generalized (i.e. $Q^2$ dependent) Gerasimov-Drell-Hearn (GDH) integral, 
as obtained in this model \cite{DKT99}. 
Indeed, at small $Q^2$, the GDH integral is 
largely dominated by the $\Delta(1232)$ resonance, whereas with increasing 
$Q^2$, the $\Delta (1232)$ contribution drops rapidly and the higher resonance 
region turns over the sign of the GDH integral, approaching its value 
as measured in deep inelastic scattering. It will be interesting to 
confirm this behavior by comparing the values of $B_n$ at $E_e = 3$~GeV 
from forthcoming data at forward and backward angles~\cite{Happex,G0}.   
\newline
\indent
We also note from Fig.~\ref{fig:int_bn4}, that at 
$E_e = 3$~GeV, the $\pi N$ contribution is only known for $W < 2$~GeV, 
whereas the upper integration range in $W$ 
is given by $W_{max} \simeq 2.55$~GeV. One can deduce from 
Fig.~\ref{fig:int_bn4} that there might be an additional negative contribution 
to $B_n$ in particular in the forward angular range. This may render 
the beam normal spin asymmetry somewhat more negative 
in the forward angular range than shown on Fig.~\ref{fig:bn_3p0}.
\begin{figure}[h]
\includegraphics[width=12cm]{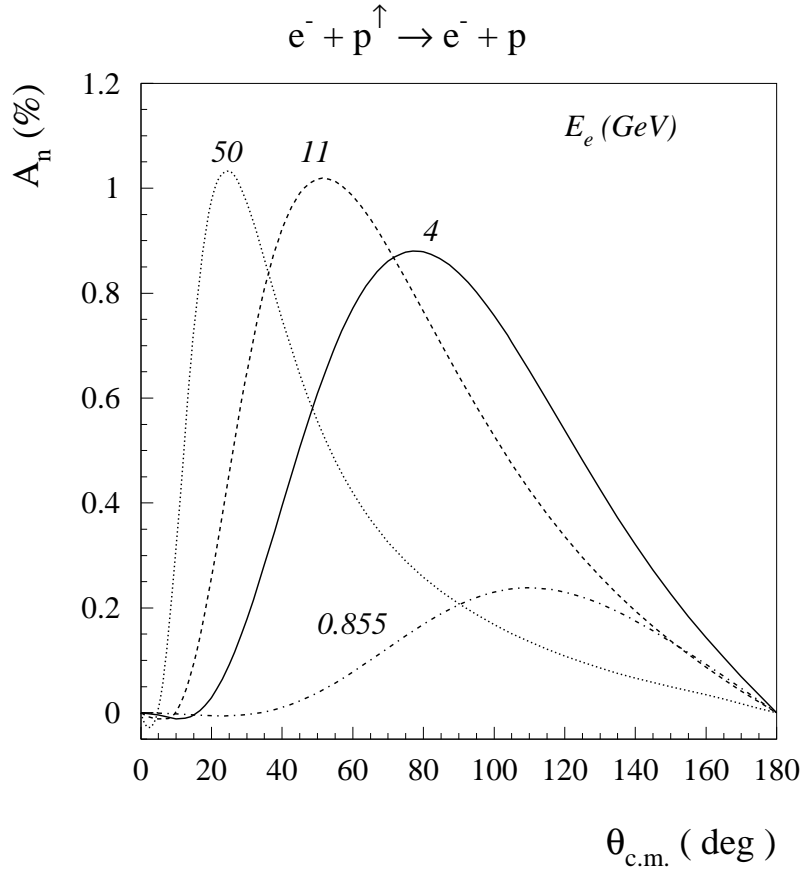}
\caption{Target normal spin asymmetry $A_n$ for $e^- p^\uparrow \to e^- p$ 
for the nucleon  
intermediate state (i.e. elastic contribution $X = N$) 
in the blob of Fig.~\ref{fig:2gamma}, as function of the c.m. scattering angle 
for different beam energies as indicated on the curves. }
\label{fig:an_elast}
\end{figure}
\newline
\indent
In the following figures, we discuss the corresponding target normal spin 
asymmetry $A_n$. We firstly show in Fig.~\ref{fig:an_elast} 
the elastic contribution to the target 
normal spin asymmetry $A_n$ at different beam energies. The elastic 
contribution to $A_n$ 
depends only on the on-shell nucleon electromagnetic form factors and has been 
calculated long time ago (see e.g. Ref.~\cite{RKR71}). Using dipole form 
factor parametrizations for both $G_{M \, p}$ and $G_{E \, p}$ 
as adopted in Ref.~\cite{RKR71}, we are able to exactly reproduce the 
results of Ref.~\cite{RKR71}. One sees from Fig.~\ref{fig:an_elast} 
that the elastic contribution to $A_n$ is around or below 1 \%. 
\begin{figure}[h]
\includegraphics[width=12cm]{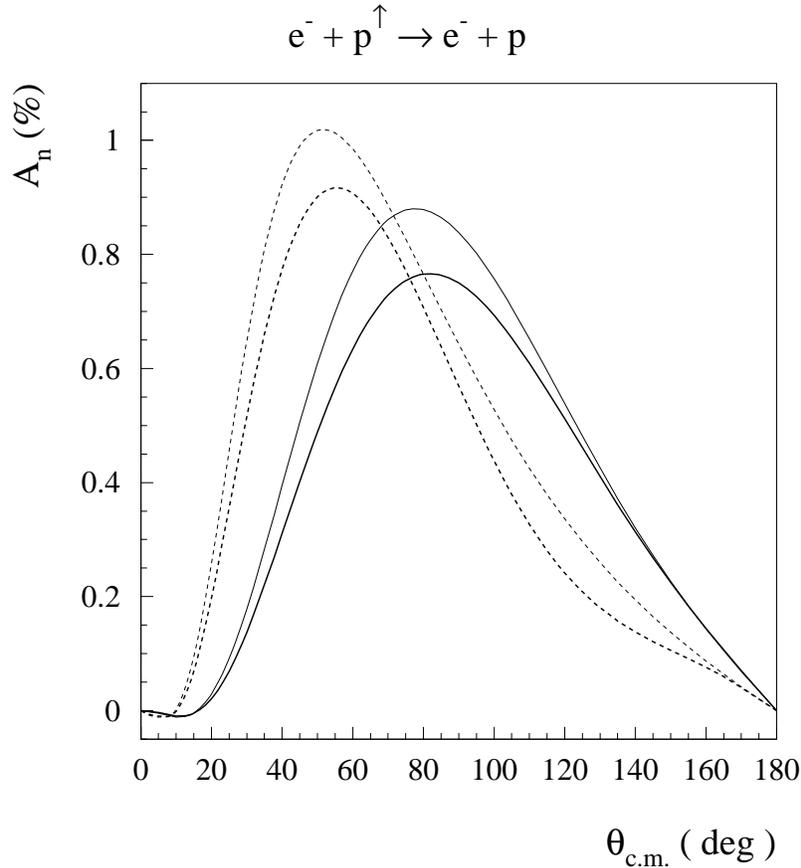}
\caption{Dependence on the proton form factors of the 
elastic contribution to the target normal spin asymmetry 
$A_n$ for $e^- p^\uparrow \to e^- p$ 
at two beam energies : $E_e = 11$~GeV (dashed curves), 
and $E_e = 4$~GeV (solid curves). 
The thin upper curves are obtained using dipole form factors. 
The thick lower curves are obtained 
using the $G_{Mp}$ parametrization of Ref.~\cite{Bra02}, 
and taking the $G_{E p} / G_{M p}$ ratio from Ref.~\cite{Gayou02}.}  
\label{fig:an_ff}
\end{figure}
\newline
\indent
In Fig.~\ref{fig:an_ff} we test the dependence of the elastic contribution 
on the on-shell proton electric and magnetic form factors. 
We compare the result for $A_n$ obtained using dipole form factors, with 
the elastic contribution calculated using the recent experimental analyses of 
$G_{M p}$ from Ref.~\cite{Bra02} and 
taking the $G_{E p} / G_{M p}$ ratio from Ref.~\cite{Gayou02}. 
One notices that the realistic form factors reduce $A_n$ by around 0.1 \% 
at its maximum. At lower beam energies (corresponding with lower values of 
$Q^2$), the deviations from the dipole parametrization of the 
form factors are much smaller. Unless otherwise stated, our results for 
the elastic contributions to the beam and target normal SSA 
are therefore calculated using dipole form factors for the proton. 
\begin{figure}[h]
\includegraphics[width=14cm]{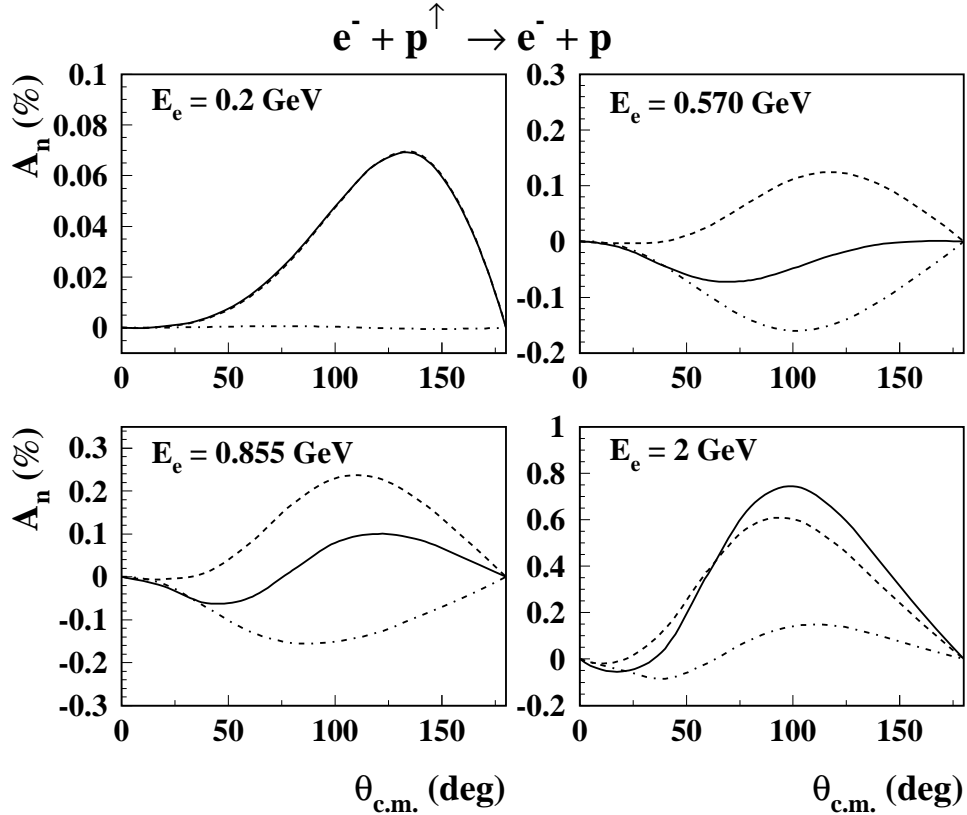}
\caption{Target normal spin asymmetry $A_n$ for $e^- p^\uparrow \to e^- p$
as function of the $c.m.$ scattering angle 
at different beam energies, as indicated on the figure.  
The calculations are for different hadronic 
intermediate states ($X$) in the blob of Fig.~\ref{fig:2gamma} ~: 
$N$ (dashed curve),  
$\pi N$  (dashed-dotted curve), 
sum of $N$ and $\pi N$ (solid curve).}
\label{fig:an_endep}
\end{figure}
\newline
\indent
In Fig.~\ref{fig:an_endep}, we show the results for 
both elastic and inelastic contributions to $A_n$ at different 
beam energies. At a low beam energy of $E_e = 0.2$~GeV, $A_n$ is 
completely dominated by the elastic contribution. Going to higher beam 
energies, the inelastic contribution becomes of comparable magnitude 
as the elastic one. This is in contrast with the situation for $B_n$ 
where the elastic contribution already becomes negligible for beam 
energies around $E_e = 0.3$~GeV. We also notice from Fig.~\ref{fig:an_endep} 
that for beam energies below 1~GeV 
the elastic and inelastic contributions to $A_n$ have opposite sign. 
The integrand of the inelastic contribution at a beam energy of 
$E_e = 0.855$~GeV is shown in Fig.~\ref{fig:int_an1}. The 
total inelastic result displays a $\pi^+ n$ threshold region contribution 
and a peak at the $\Delta(1232)$ resonance. Notice that the higher resonance 
region is suppressed in comparison with the corresponding integrand 
for $B_n$. Also the quasi-real Compton scattering peak around the maximum  
$W$ value is absent. This different behavior 
in comparison with the beam normal spin asymmetry can be easily understood 
by comparing the lepton tensors in both cases. 
One sees from Eq.~(\ref{eq:lept1}) that the unpolarized lepton tensor, 
which enters in $A_n$, vanishes linearly when the intermediate lepton momentum 
$k_1 \to 0$. This is 
in contrast with the polarized lepton tensor of Eq.~(\ref{eq:lept3}) 
which becomes constant when $k_1 \to 0$. 
Hence the region around $W = W_{max}$ (corresponding with $k_1 \to 0$) 
in the integrand of $A_n$ is suppressed compared with the corresponding 
region in the integrand of $B_n$. As a result, the elastic contribution 
to $A_n$ can be of comparable magnitude as the inelastic contribution.   
Furthermore, one sees from Fig.~\ref{fig:an_endep} that, due to the partial 
cancellation between elastic and inelastic contributions, $A_n$ 
for the proton is significantly reduced, 
taking on values around or below 0.1 \% for beam energies below 1~GeV. 
\begin{figure}[h]
\includegraphics[width=14cm]{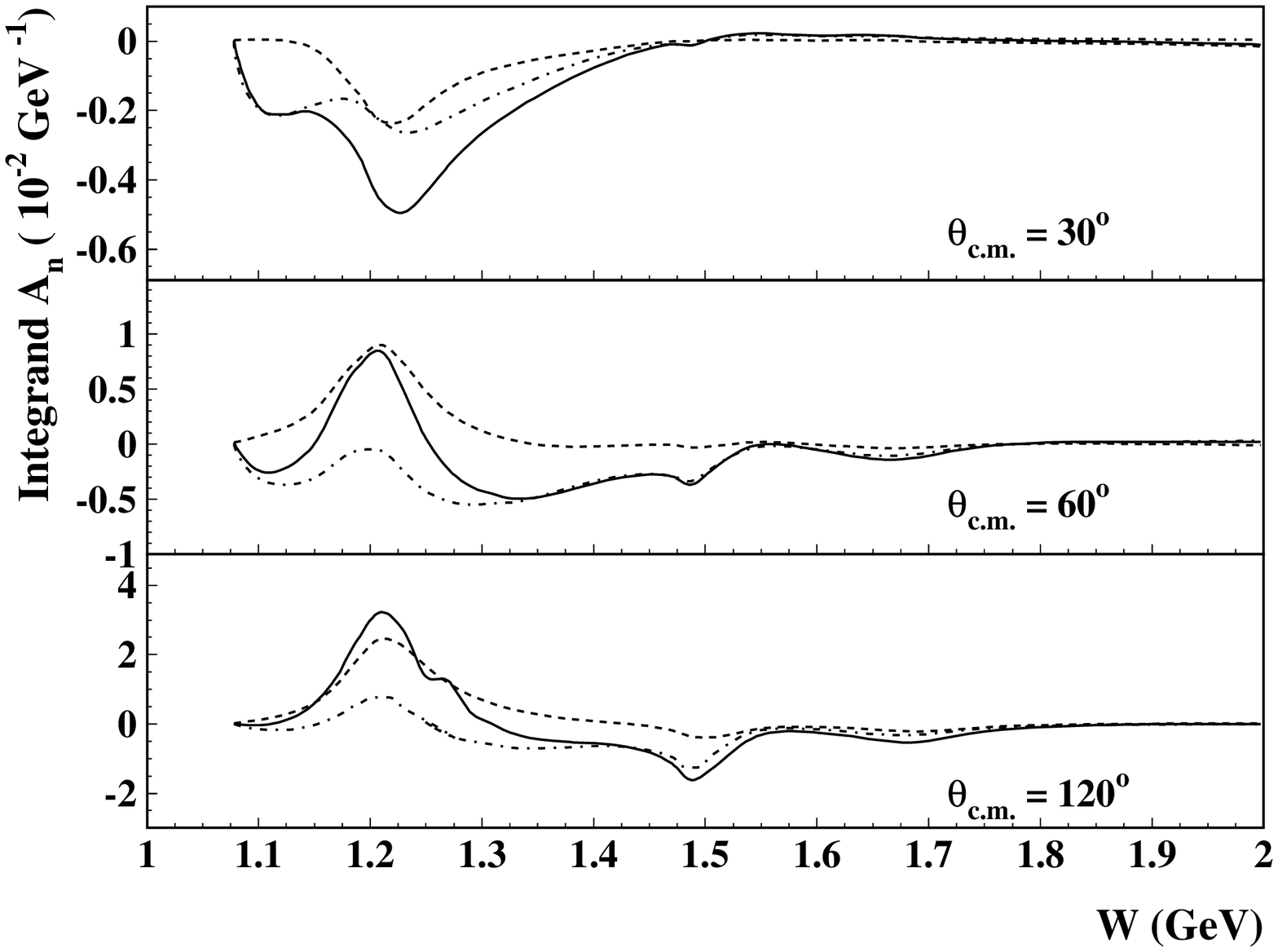}
\vspace{-1cm}
\caption{
Integrand in $W$ of the target normal spin asymmetry $A_n$ 
for $e^- p^\uparrow \to e^- p$
at a beam energy of $E_e = 2$ GeV and at different $c.m.$ scattering 
angles as indicated on the figure. 
The dashed curves are the contribution from the $\pi^0 p$ channel,
the dashed-dotted curves show the contribution from the $\pi^+ n$ 
channel, and the solid curves are the sum of the  
$\pi^+ n$ and $\pi^0 p$ channels.}
\label{fig:int_an2}
\end{figure}
\newline
\indent
At higher beam energies, the inelastic contribution to $A_n$ changes 
sign. This can be understood by comparing the integrands of $A_n$ 
at $E_e = 0.855$~GeV (Fig.~\ref{fig:int_an1}) with its value 
at $E_e = 2$~GeV (Fig.~\ref{fig:int_an2}). One sees that at $E_e = 2$~GeV  
and backward angles, the $\Delta(1232)$ contribution changes sign and 
dominates the inelastic contribution. Because at higher energies also 
the elastic contribution grows larger, as was seen in Fig.~\ref{fig:an_elast}, 
one obtains larger target normal spin asymmetries around 1 \%.  
\begin{figure}
\includegraphics[width=12cm]{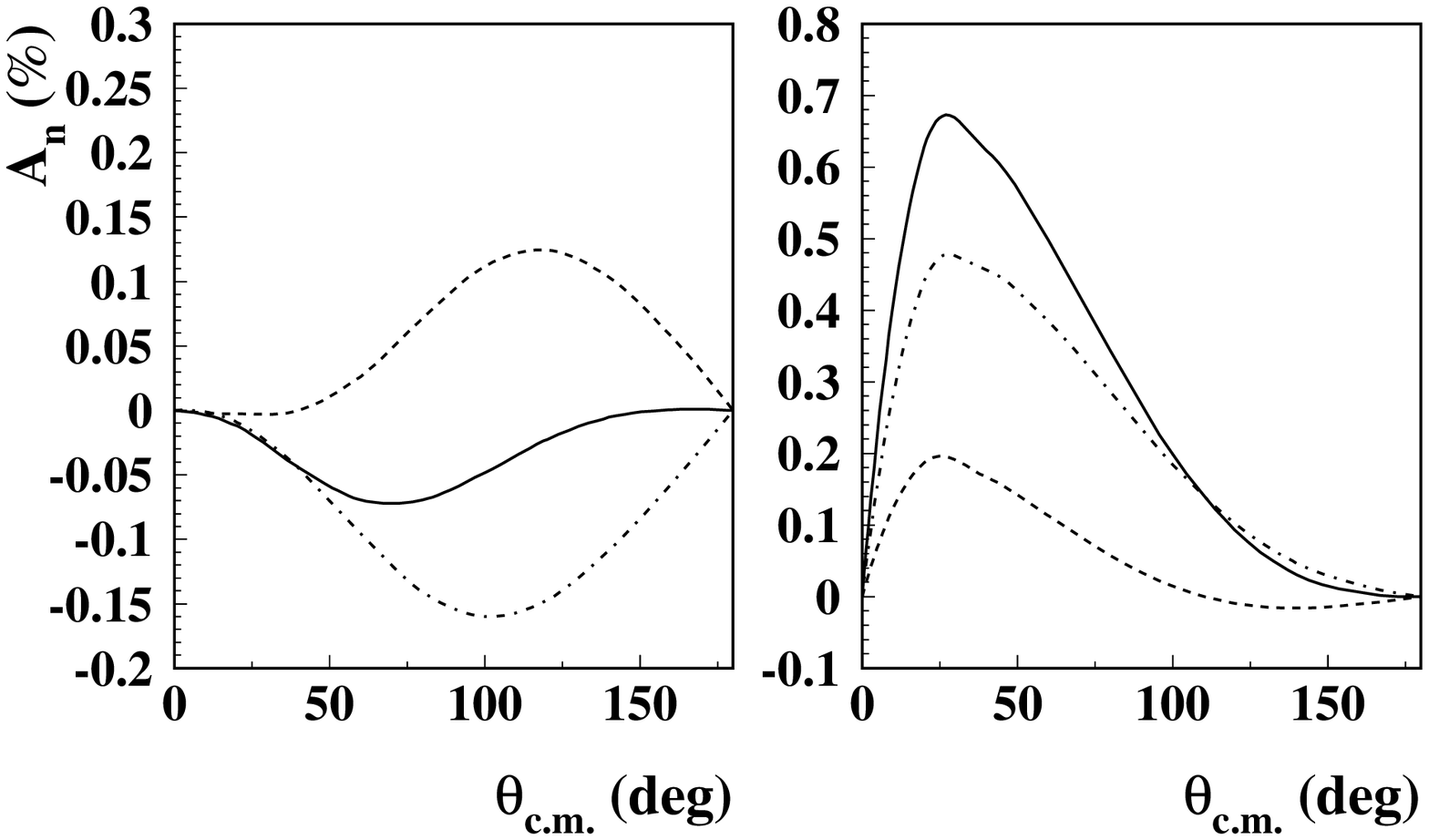}
\caption{Comparison of the target normal spin asymmetry $A_n$ for the processes
$e^- p^\uparrow \to e^- p$ (left panel), and   
$e^- n^\uparrow \to e^- n$ (right panel) 
at beam energy $E_e=0.570$ GeV
as function of the $c.m.$ scattering angle,  
for different hadronic intermediate states ($X$) in the blob  of
Fig.~\ref{fig:2gamma} ~: 
$N$ (dashed curve),  
$\pi N$  (dashed-dotted curve), 
sum of $N$ and $\pi N$ (solid curve).}
\label{fig:tnsa_neutron}
\end{figure}
\newline
\indent
In Fig.~\ref{fig:tnsa_neutron}, we compare the target normal spin asymmetries 
for elastic electron scattering off protons and neutrons. 
For the elastic contribution to $A_n$ for the neutron, we use 
the parametrizations of 
$G_{M n}$ from Ref.~\cite{Kub02}, and $G_{E n}$ from Ref.~\cite{War04}. 
The inelastic 
contribution to $A_n$ for the neutron is enhanced in comparison 
with the proton. This can be understood from Eq.~(\ref{eq:tnsa}) for $A_n$. 
For the proton, both $G_M$ and $G_E$ terms are sizeable and tend to 
cancel each other. For the neutron on the other hand, 
the $G_M$ term changes sign whereas the $G_E$ term is very small 
so that such cancellation does not occur. Therefore, 
the target normal spin asymmetry is quite sizeable for the neutron 
(around 0.65 \%) in the resonance region, providing an interesting 
opportunity for a measurement. 

\section{Conclusions}
\label{sec:concl}

In this paper, we have studied the target and beam normal single spin 
asymmetries for elastic electron-nucleon scattering. 
These asymmetries depend on the imaginary part of $2 \gamma$ exchange 
amplitudes. We have constructed the imaginary part of these 
$2 \gamma$ exchange amplitudes as a phase space integral over the doubly 
virtual Compton scattering tensor on the nucleon. 
Using unitarity, we have expressed the imaginary (absorptive) part of the 
non-forward doubly virtual Compton tensor on the nucleon 
in the resonance region in terms of phenomenological $\gamma^* N \to \pi N$ 
electroproduction amplitudes. 
\newline
\indent
Using this model for the non-forward doubly 
virtual Compton tensor, we presented calculations for beam and target normal 
SSAs for several experiments performed or in progress. 
The resonance region, where the model 
input is relatively well understood, is a useful testing ground 
to study these asymmetries as a new tool to extract nucleon 
structure information. 
\newline
\indent
At a low beam energy, around pion threshold, 
the inelastic ($\pi N$ intermediate state) contribution is largely constrained 
from chiral symmetry predictions. Around pion threshold,
the beam normal SSA $B_n$ is at the few ppm level. 
Going up in beam energy, the elastic contribution to $B_n$ becomes very soon 
negligible (at the 1 ppm level) whereas the resonance contributions yield 
large values of $B_n$ of the order of several tens 
of ppm in the backward angular range. 
This is mainly driven by the quasi-VCS and quasi-RCS near singularities, in 
which one or both photons in the two-photon exchange process 
become quasi-real. 
It was found that at forward angles, the size of the predicted asymmetries 
is compatible with the first high precision measurements performed at 
MAMI. It will be interesting to check that for backward angles the 
beam normal SSA indeed grows to the level of tens of ppm in the 
resonance region. 
\newline
\indent
For higher beam energies, around 3 GeV energy range, 
the inelastic contribution to $B_n$ was found to display 
an interesting structure :  
it is negative (around - 3 ppm) in the forward angular range, and changes 
sign around $\theta_{c.m.} \simeq 90^o$. This behavior can be 
understood from the observation that at forward angles the three 
main resonance regions enter with the same sign in $B_n$. At backward angles 
however, the first two resonance regions are largely damped and the 
third resonance region drives the change of sign in $B_n$. 
\newline
\indent
We have also shown our results for the target normal SSA $A_n$. 
In contrast to the beam normal SSA, the quasi-RCS near 
singularity is absent in the target normal SSA, yielding much 
smaller inelastic contributions relative to elastic ones. 
At beam energies around 1 GeV, elastic and inelastic contributions to $A_n$ 
tend to cancel each other for the proton, yielding values for $A_n$ around 
0.1 \%. For the neutron, such a cancellation is absent and one may expect 
values of $A_n$ approaching 1 \%. 
\newline
\indent
Besides providing estimates for ongoing experiments, this work 
can be considered as a first step in the construction of a 
dispersion formalism for elastic electron-nucleon scattering amplitudes. 
In such a formalism, one needs a precise knowledge of the imaginary part 
as input in order to construct the real part as a dispersion integral 
over this imaginary part. The   
real part of the two-photon exchange amplitudes may yield  
corrections to elastic electron-nucleon scattering observables, 
such as the unpolarized cross sections or double polarization observables. 
It is of importance to quantify this piece of information, in order 
to increase the precision in the extraction of nucleon form factors. 
\newline
\indent
Besides, this work may also be extended to the calculation of 
$\gamma Z$ and $W^+ W^-$ box diagrams, which enter as corrections in 
electroweak precision experiments.

\section*{Acknowledgments}

This work was supported 
by the Deutsche Forschungsgemeinschaft (SFB443), 
by the Italian MIUR through the PRIN Theoretical Physics 
of the Nucleus and the Many-Body Systems,  
and by the U.S. Department of Energy under contracts 
DE-FG02-04ER41302 and DE-AC05-84ER40150. 
The authors also thank the Institute for Nuclear 
Theory at the University of Washington  and the ECT* in Trento, 
where part of this work was performed, for their hospitality. 
Furthermore, the authors thank A. Afanasev, 
C. Carlson, M. Gorchtein, P.A.M. Guichon, F. Maas, 
and S. Wells for helpful discussions.

\begin{appendix}

\section{Relations between helicity amplitudes and invariant
amplitudes for elastic electron-nucleon scattering}
\label{app:hel}

The helicity amplitudes for elastic electron-nucleon scattering are
defined in the $e N$ $c.m.$ frame, and are denoted by  
$T(h' , \lambda_N' \,;\, h ,\lambda_N)$, 
where $h (h')$ denote the helicities of the initial (final) electrons
and 
where $\lambda_N (\lambda_N')$ denote the helicities of the 
initial (final) nucleons. It is also convenient to introduce the 
Mandelstam invariants $s = (p + k)^2$ and $u = (p - k')^2$ which, 
neglecting the electron mass, are related to the
invariants $Q^2$ and $\nu$, introduced in Eq.~(\ref{eq:qsqrnu}), as~:
\begin{equation}
s + u = Q^2 + 2 \, M^2, \hspace{1cm} s - u = 4 \, \nu.
\end{equation}
Furthermore, the $c.m.$ scattering angle
$\theta_{c.m.}$ is related to $s, u$, and $Q^2$ as~:
\begin{equation}
\sin^2 \frac{\theta_{c.m.}}{2} \,=\, \frac{Q^2 \, s}{(s - M^2)^2} ,
\hspace{1cm}
\cos^2 \frac{\theta_{c.m.}}{2} \,=\, \frac{(M^4 - s u)}{(s - M^2)^2} .
\end{equation}
\indent
The helicity spinors for the electrons are given by~:
\beqn
u(k(k'), h(h'))\;=\;\sqrt{|\vec k|}
\left[
\begin{array}{c}
\chi_{h(h')}\\
2h(2h') \; \chi_{h(h')}
\end{array}
\right],
\eeqn
where $| \vec k| = (s - M^2) / (2 \sqrt{s})$, and   
where the Pauli spinors for the incoming electron are given by~:
\beqn
\chi_{1\over 2}\;=\;\left(\begin{array}{c}1\\0\end{array}\right)
\;\;\;\;\;\;\;\;
\chi_{-{1\over 2}}\;=\;\left(\begin{array}{c}0\\1\end{array}\right)\;,
\eeqn
whereas the Pauli spinors for the outgoing electron are given by~:
\beqn
\chi'_{1\over 2}\;=\;
\left(
\begin{array}{c}
\cos{{\theta_{c.m.}}\over 2}\\
\sin{{\theta_{c.m.}}\over 2}
\end{array}
\right),
\;\;\;\;\;\;\;\;
\chi'_{-{1\over 2}}\;=\;
\left(
\begin{array}{c}
-\sin{{\theta_{c.m.}}\over 2}\\
\cos{{\theta_{c.m.}}\over 2}
\end{array}
\right).
\eeqn
The helicity spinors for the nucleon are given by~:
\beqn
u(p(p'), \lambda_N(\lambda_N'))\;=\;\sqrt{E_N+M}
\left[
\begin{array}{c}
\chi_{\lambda_N (\lambda_N')}\\
2 \lambda_N (2\lambda_N') \;\frac{|\vec k|}{E_N+M}\;
\chi_{\lambda_N(\lambda_N')}
\end{array}
\right],
\eeqn
where $E_N = \sqrt{|\vec k|^2 + M^2}$. 
The Pauli spinors for the initial proton are given by~:
\beqn
\chi_{1\over 2}\;=\;\left(\begin{array}{c}0\\-1\end{array}\right)
\;\;\;\;\;\;\;\;
\chi_{-{1\over 2}}\;=\;\left(\begin{array}{c}1\\0\end{array}\right)\;,
\eeqn
and the Pauli spinors for the final proton are given by~:
\beqn
\chi'_{1\over 2}\;=\;
\left(
\begin{array}{c}
\sin{{\theta_{c.m.}}\over 2}\\
-\cos{{\theta_{c.m.}}\over 2}
\end{array}
\right)
\;\;\;\;\;\;\;\;
\chi'_{-{1\over 2}}\;=\;
\left(
\begin{array}{c}
\cos{{\theta_{c.m.}}\over 2}\\
\sin{{\theta_{c.m.}}\over 2}
\end{array}
\right)\;.
\eeqn
\newline
\indent
Using the constraints of parity invariance and time reversal
invariance, one obtains 3 independent helicity amplitudes which conserve the
electron helicity (i.e. $h' = h$), and 3 independent helicity amplitudes
which flip the electron helicity (i.e. $h' = - h$), in
agreement with the invariants found in
Eqs.~(\ref{eq:non-flip},\ref{eq:flip}).
\newline
\indent
In terms of the invariants $\tilde G_M$, $\tilde F_2$, and $\tilde
F_3$, the 3 independent 
helicity amplitudes which conserve the electron helicity can be expressed as~:
\begin{eqnarray}
T_1 \,&\equiv& \, T(h'= +\frac{1}{2} , \lambda_N' = +\frac{1}{2} \, ;\, 
h = +\frac{1}{2} , \lambda_N = +\frac{1}{2}) \nonumber \\
&=& - \frac{e^2}{Q^2} \, (s - M^2) \, 
\left\{ - 2 \, \tilde G_M \, \left[ \frac{M^2}{s} 
\cos^2 \frac{\theta_{c.m.}}{2}
\,+\, \frac{(s - M^2)}{s} \right] \right. \nonumber\\ 
&&\left. \hspace{3cm} +\, 2 \, \tilde F_2 \, \cos^2 \frac{\theta_{c.m.}}{2}
\,-\, \tilde F_3 \, \frac{(s - M^2)}{M^2} \, \cos^2 \frac{\theta_{c.m.}}{2} 
\right\}, 
\label{eq:t1} \\ 
&& \nonumber\\
T_2 \, &\equiv& \, T(h'= +\frac{1}{2} , \lambda_N' = -\frac{1}{2} \, ;\, 
h = +\frac{1}{2} , \lambda_N = +\frac{1}{2}) \nonumber \\
&=& - \frac{e^2}{Q^2} \, \frac{(s - M^2)}{M \, \sqrt{s}} \,  
\sin \frac{\theta_{c.m.}}{2} \, \cos \frac{\theta_{c.m.}}{2} 
 \left\{ \tilde G_M \, (2 \, M^2) \,-\, 
\tilde F_2 \,(s + M^2) \,+\, \tilde F_3 \,(s - M^2) \right\} , 
\nonumber\\
 & &\label{eq:t2} \\
T_3 \, &\equiv& \, T(h'= +\frac{1}{2} , \lambda_N' = -\frac{1}{2} \, ;\, 
h = +\frac{1}{2} , \lambda_N = -\frac{1}{2}) \nonumber \\
&=& - \frac{e^2}{Q^2} \, (s - M^2) \,  
\cos^2 \frac{\theta_{c.m.}}{2} 
\, \left\{ - 2 \,  \tilde G_M  \,+\, 
2 \, \tilde F_2 \,-\, \tilde F_3 \, \frac{(s - M^2)}{M^2} \right\} .
\label{eq:t3}
\end{eqnarray}
Inverting the relations in Eqs.~(\ref{eq:t1} - \ref{eq:t3}), yield the 
invariant amplitudes $\tilde G_M$, $\tilde F_2$, and $\tilde F_3$ as~:
\begin{eqnarray}
e^2 \, \tilde G_M \,&=&\,  \frac{1}{2} \, \left\{ \, T_1 - T_3 \, \right\} , 
\label{eq:gmhel} \\
e^2 \, \tilde F_2 \,&=&\,  \frac{M}{\sqrt{s}} \, \tan \frac{\theta_{c.m.}}{2} 
\, \left\{ \, T_2 \,+\, \frac{M}{\sqrt{s}} \, \tan \frac{\theta_{c.m.}}{2} \, 
T_3 \, \right\} , 
\label{eq:f2thel} \\
e^2 \, \tilde F_3 \,&=&\,  \frac{M^2}{(s - M^2)} \,
\left\{ \, - T_1 \,+\, \frac{2 M}{\sqrt{s}} \, \tan \frac{\theta_{c.m.}}{2} \, 
T_2 \,+\, \left( 1 + \frac{s + M^2}{s} 
\tan^2 \frac{\theta_{c.m.}}{2} \right) T_3 \, \right\} .
\nonumber\\
& &\label{eq:f3thel} 
\end{eqnarray}
\indent
The three amplitudes which flip the electron helicity can be expressed as~:
\beqn
T_4&\equiv&
T(h'=-\frac{1}{2},\lambda'_N=\frac{1}{2};h=\frac{1}{2},\lambda_N=\frac{1}{2})
\nn\\
&=&- \frac{m_e \, e^2}{2 \, k} \frac{1}{\tan{{\theta_{c.m.}}\over 2}}
\left[\frac{2 (s + M^2)}{(s - M^2)}(\tilde{G}_M-\tilde{F}_2)
+\frac{s+M^2}{M^2}\tilde{F}_3
\,+\,2 \tilde{F}_4\,+\,\frac{s-M^2}{M^2}\tilde{F}_5
\right] , \\
T_5&\equiv&
T(h'=-\frac{1}{2},\lambda'_N=-\frac{1}{2};h=\frac{1}{2},\lambda_N=\frac{1}{2})
\nn\\
&=&\frac{m_e \, e^2}{M}
\left[
\frac{4 \, s \, M^2}{(s - M^2)^2} \tilde{G}_M 
\,-\, \frac{(s + M^2)^2}{(s - M^2)^2} \tilde{F}_2
\,+\, \frac{(s + M^2)}{(s - M^2)} (\tilde{F}_3+\tilde{F}_4)
+ \tilde{F}_5 + \tilde{F}_6
\right] ,\\
T_6&\equiv&
T(h'=-\frac{1}{2},\lambda'_N=\frac{1}{2};h=\frac{1}{2},\lambda_N=-\frac{1}{2})
\nn\\
&=&-\frac{m_e \, e^2}{M}
\left[
\frac{4 \, s \, M^2}{(s - M^2)^2}\tilde{G}_M
\,-\, \frac{(s + M^2)^2}{(s - M^2)^2} \tilde{F}_2
\,+\, \frac{(s + M^2)}{(s - M^2)} (\tilde{F}_3+\tilde{F}_4)
+ \tilde{F}_5 - \tilde{F}_6
\right] .
\nonumber
\\
& &
\eeqn
The inversion of these relations reads~:
\beqn
e^2\tilde{F}_4&=&-\frac{M^2}{2\sqrt{s}k}
\left[
T_1\,-\,\frac{(s + M^2)}{\sqrt{s} \, M}\tan{{\theta_{c.m.}}\over 2}T_2\,-\,
\left(1+\frac{(s + M^2)}{s}\tan^2{{\theta_{c.m.}}\over 2}\right)T_3
\right] \nonumber \\
&-&\frac{M}{2m_e}(T_6-T_5)
\;+\;\frac{M^2}{\sqrt{s} \, m_e}\tan{{\theta_{c.m.}}\over 2}T_4 \\
e^2\tilde{F}_5&=&-\frac{M^4}{2sk^2}
\left[
-T_1\,+\,\frac{(s + M^2)}{\sqrt{s} \, M}\tan{{\theta_{c.m.}}\over 2}T_2
\,+\,\left(1+\frac{(s + M^2)^2}{2 \, s \, M^2}\tan^2{{\theta_{c.m.}}\over 2}\right)T_3
\right] \nonumber \\
&+&\frac{M^3}{2m_e\sqrt{s}k}(T_6-T_5)
\;-\;\frac{M^2 (s + M^2)}{m_e\, 2 \, s \, k}
\tan{{\theta_{c.m.}} \over 2}T_4  \\
e^2\tilde{F}_6&=&\frac{M}{2m_e}(T_5+T_6)\; .
\eeqn

\end{appendix}

\end{document}